\documentclass[%
reprint,
superscriptaddress,
 amsmath,amssymb,
aps,
 prb,
floatfix,
]{revtex4-2}

\usepackage[citecolor=blue, urlcolor=blue, linkcolor=blue, colorlinks=true]{hyperref}

\usepackage[ruled,vlined]{algorithm2e}
\usepackage{enumitem}
\usepackage{graphicx}%
\usepackage{dcolumn}%
\usepackage{bm}%

\usepackage{braket}
\usepackage{bbold}
\usepackage{diagbox}
\usepackage[table, x11names, svgnames]{xcolor}

\usepackage{xcolor}
\usepackage{soul}

\newcommand{\shenghide}[1] {}

\newcommand{\rewrite}[2] {#2} %

\DeclareMathOperator*{\argmax}{arg\,max\ }
\DeclareMathOperator*{\argmin}{arg\,min\ }

\usepackage{tikz, ifthen}
\usetikzlibrary{decorations.markings}
\usetikzlibrary{decorations.pathmorphing}

\newenvironment{diagram}
{
\begin{tikzpicture}[baseline = (X.base),every node/.style={scale=0.7},scale=.55]
}
{
\end{tikzpicture}
}

\newcommand{\drawMatrixLeft}[1]{
\begin{diagram}
draw (0,0) node (X) {};  %
\draw (-0.5, 0.) edge[out=90,in=180] (1.,1.75);
\draw (-0.5, -0.) edge[out=270,in=180] (1.,-1.75);
\draw (-0., 0.) edge[out=90,in=180] (1.,1.25);
\draw (-0., -0.) edge[out=270,in=180] (1.,-1.25);
\end{diagram}
}
\newcommand{\drawMatrixRight}[1]{
\begin{diagram}
\draw (0,0) node (X) {};  %
\draw (1, 0) edge[out = -90, in = 0] (0, -1.25);
\draw (1, 0) edge[out = 90, in = 0] (0, 1.25);
\draw (1.5, 0) edge[out = -90, in = 0] (0, -1.75);
\draw (1.5, 0) edge[out = 90, in = 0] (0, 1.75);
\end{diagram}
}
\newcommand{\applyTransferRight}[3]{
\begin{diagram}
\draw (0,0) node (X) {};  %
\draw (4, 0) edge[out = -90, in = 0] (3, -1.25);
\draw (4, 0) edge[out = 90, in = 0] (3, 1.25);
\draw (4.5, 0) edge[out = -90, in = 0] (3, -1.75);
\draw (4.5, 0) edge[out = 90, in = 0] (3, 1.75);
\draw[rounded corners] (1,2.5) rectangle (3,0.5);
\draw[rounded corners] (1,-0.5) rectangle (3,-2.5);
\draw (1,-1.75) -- (3.,-1.75);
\draw (1,1.75) -- (3.,1.75);
\draw (1.75,0.5) -- (1.75,-0.5);
\draw (2.25,0.5) -- (2.25,-0.5);
\draw (1,1.25) edge[out=0,in=90] (1.75,0.5);
\draw (1,-1.25) edge[out =0, in=-90] (1.75, -0.5);
\draw (3,1.25) edge[out =180, in=90] (2.25, 0.5);
\draw (3,-1.25) edge[out=180, in=270] (2.25, -0.5);
\draw (0.5,-1.25) -- (1,-1.25);
\draw (0.5, 1.25) -- (1, 1.25);
\draw (0.5,-1.75) -- (1,-1.75);
\draw (0.5, 1.75) -- (1, 1.75);

\end{diagram}
}
\newcommand{\applyTransferLeft}[3]{
\begin{diagram}
\draw (0,0) node (X) {};  %
\draw (-0.5, 0.) edge[out=90,in=180] (1.,1.75);
\draw (-0.5, -0.) edge[out=270,in=180] (1.,-1.75);
\draw (-0., 0.) edge[out=90,in=180] (1.,1.25);
\draw (-0., -0.) edge[out=270,in=180] (1.,-1.25);
\draw[rounded corners] (1,2.5) rectangle (3,0.5);
\draw[rounded corners] (1,-0.5) rectangle (3,-2.5);
\draw (1,-1.75) -- (3.5,-1.75);
\draw (1,1.75) -- (3.5,1.75);
\draw (1.75,0.5) -- (1.75,-0.5);
\draw (2.25,0.5) -- (2.25,-0.5);
\draw (1,1.25) edge[out=0,in=90] (1.75,0.5);
\draw (1,-1.25) edge[out =0, in=-90] (1.75, -0.5);
\draw (3,1.25) edge[out =180, in=90] (2.25, 0.5);
\draw (3,-1.25) edge[out=180, in=270] (2.25, -0.5);
\draw (3.,-1.25) -- (3.5,-1.25);
\draw (3., 1.25) -- (3.5, 1.25);
\end{diagram}
}

\newcommand{\drawMatrixLeftIso}[1]{
\begin{diagram}
draw (0,0) node (X) {};  %
\draw (-0., 0.) edge[out=90,in=180] (1.,1.25);
\draw (-0., -0.) edge[out=270,in=180] (1.,-1.25);
\end{diagram}
}
\newcommand{\drawMatrixRightIso}[1]{
\begin{diagram}
\draw (0,0) node (X) {};  %
\draw (1, 0) edge[out = -90, in = 0] (0, -1.25);
\draw (1, 0) edge[out = 90, in = 0] (0, 1.25);
\end{diagram}
}
\newcommand{\applyTransferRightIso}[3]{
\begin{diagram}
\draw (0,0) node (X) {};  %
\draw (4, 0) edge[out = -90, in = 0] (3, -1.25);
\draw (4, 0) edge[out = 90, in = 0] (3, 1.25);
\draw[rounded corners] (1,2.5) rectangle (3,0.5);
\draw[rounded corners] (1,-0.5) rectangle (3,-2.5);
\draw (1.75,0.5) -- (1.75,-0.5);
\draw (2.25,0.5) -- (2.25,-0.5);
\draw (1,1.25) edge[out=0,in=90] (1.75,0.5);
\draw (1,-1.25) edge[out =0, in=-90] (1.75, -0.5);
\draw (3,1.25) edge[out =180, in=90] (2.25, 0.5);
\draw (3,-1.25) edge[out=180, in=270] (2.25, -0.5);
\draw (0.5,-1.25) -- (1,-1.25);
\draw (0.5, 1.25) -- (1, 1.25);
\end{diagram}
}
\newcommand{\applyTransferLeftIso}[3]{
\begin{diagram}
\draw (0,0) node (X) {};  %
\draw (-0., 0.) edge[out=90,in=180] (1.,1.25);
\draw (-0., -0.) edge[out=270,in=180] (1.,-1.25);
\draw[rounded corners] (1,2.5) rectangle (3,0.5);
\draw[rounded corners] (1,-0.5) rectangle (3,-2.5);
\draw (1.75,0.5) -- (1.75,-0.5);
\draw (2.25,0.5) -- (2.25,-0.5);
\draw (1,1.25) edge[out=0,in=90] (1.75,0.5);
\draw (1,-1.25) edge[out =0, in=-90] (1.75, -0.5);
\draw (3,1.25) edge[out =180, in=90] (2.25, 0.5);
\draw (3,-1.25) edge[out=180, in=270] (2.25, -0.5);
\draw (3.,-1.25) -- (3.5,-1.25);
\draw (3., 1.25) -- (3.5, 1.25);
\end{diagram}
}
\newcommand{\drawMatrixMixedIso}[1]{
\begin{diagram}
\draw (0,0) node (X) {};  %
\draw (1, 0) edge[out = -90, in = 0] (0, -1.25);
\draw (1, 0) edge[out = 90, in = 0] (0, 1.25);
\draw (1.5-0., 0.) edge[out=90,in=180] (1.5+1.,1.25);
\draw (1.5-0., -0.) edge[out=270,in=180] (1.5+1.,-1.25);
\end{diagram}
}
\newcommand{\applyTransferMixedIso}[3]{
\begin{diagram}
\draw (0,0) node (X) {};  %
\draw (0.5, 1.25) -- (1.,1.25);
\draw (0.5, -1.25) -- (1.,-1.25);
\draw[rounded corners] (1,2.5) rectangle (3,0.5);
\draw[rounded corners] (1,-0.5) rectangle (3,-2.5);
\draw (1.75,0.5) -- (1.75,-0.5);
\draw (2.25,0.5) -- (2.25,-0.5);
\draw (1,1.25) edge[out=0,in=90] (1.75,0.5);
\draw (1,-1.25) edge[out =0, in=-90] (1.75, -0.5);
\draw (3,1.25) edge[out =180, in=90] (2.25, 0.5);
\draw (3,-1.25) edge[out=180, in=270] (2.25, -0.5);
\draw (3.,-1.25) -- (3.5,-1.25);
\draw (3., 1.25) -- (3.5, 1.25);
\end{diagram}
}
\newcommand{\applyTransferCenterIso}[3]{
\begin{diagram}
\draw (0,0) node (X) {};  %
\draw (-0., 0.) edge[out=90,in=180] (1.,1.25);
\draw (-0., -0.) edge[out=270,in=180] (1.,-1.25);
\draw[rounded corners] (1,2.5) rectangle (3,0.5);
\draw[rounded corners] (1,-0.5) rectangle (3,-2.5);
\draw (1.75,0.5) -- (1.75,-0.5);
\draw (2.25,0.5) -- (2.25,-0.5);
\draw (1,1.25) edge[out=0,in=90] (1.75,0.5);
\draw (1,-1.25) edge[out =0, in=-90] (1.75, -0.5);
\draw (3,1.25) edge[out =180, in=90] (2.25, 0.5);
\draw (3,-1.25) edge[out=180, in=270] (2.25, -0.5);
\draw (4, 0) edge[out = -90, in = 0] (3, -1.25);
\draw (4, 0) edge[out = 90, in = 0] (3, 1.25);
\end{diagram}
}

\tikzset{->-/.style={decoration={
  markings,
  mark=at position #1 with {\arrow{latex}}},postaction={decorate}}}
\tikzset{-<-/.style={decoration={
  markings,
  mark=at position #1 with {\arrow{latex reversed}}},postaction={decorate}}}

\begin{document}

\preprint{APS/123-QED}

\title{Efficient Simulation of Dynamics in Two-Dimensional Quantum Spin Systems with Isometric Tensor Networks}%

\author{Sheng-Hsuan Lin}
\affiliation{Department of Physics, T42, Technische Universit{\"a}t M{\"u}nchen, James-Franck-Stra{\ss}e 1, D-85748 Garching, Germany}
\author{Michael P. Zaletel}%
\affiliation{
Department of Physics, University of California, Berkeley, California 94720, USA
}%

\author{Frank Pollmann}
\affiliation{Department of Physics, T42, Technische Universit{\"a}t M{\"u}nchen, James-Franck-Stra{\ss}e 1, D-85748 Garching, Germany}

\date{\today}%

\begin{abstract}
We investigate the computational power of the recently introduced class of isometric tensor network states (isoTNSs), which generalizes the isometric conditions of the canonical form of one-dimensional matrix-product states to tensor networks in higher dimensions. 
We discuss several technical details regarding the implementation of isoTNSs-based algorithms and compare different disentanglers---which are essential for an efficient handling of isoTNSs.
We then revisit the time evolving block decimation for isoTNSs ($\text{TEBD}^2$) and explore its power for real time evolution of two-dimensional (2D) lattice systems. 
Moreover, we introduce a density matrix renormalization group algorithm for isoTNSs ($\text{DMRG}^2$) that allows to variationally find ground states of 2D lattice systems. 
As a demonstration and benchmark, we compute the dynamical spin structure factor of 2D quantum spin systems for two paradigmatic models: 
First, we compare our results for the transverse field Ising model on a square lattice with the prediction of the spin-wave theory. 
Second, we consider the Kitaev model on the honeycomb lattice and compare it to the result from the exact solution.
\end{abstract}

\maketitle

\section{\label{sec:Intro} Introduction}

Quantum many-body systems are well known for their rich emergent behavior that can arise due to the interactions between large numbers of degrees of freedom. 
Prominent examples are quantum spin liquids that give rise to fractionalized excitations~\cite{stormer1999fractional,laughlin1983anomalous} and high-$T_C$ superconductivity~\cite{bednorz1986possible}.
Exact numerical simulations of such systems have generically an exponential complexity scaling in system size due to the growth of the Hilbert space dimension. 
The question to which extent an efficient solution is possible can in many cases be related to the amount of many-body entanglement~\cite{Schuch08,verstraete2008matrix,schollwock2011density,bridgeman2017hand}.
Tensor network states (TNSs) are a ``natural" way to describe quantum states with low entanglement and various algorithms have been developed to solve such problems:
For one-dimensional (1D) systems, it is known that ground states of gapped local Hamiltonian fulfills the area law~\cite{hastings2007area,eisert2010colloquium}.
This in turn tells us that matrix-product states (MPSs)~\cite{fannes1992abundance,perez2006matrix} are good approximations of those ground states~\cite{verstraete2006matrix}.
This intuition can at least partially be carried over to two-dimensional (2D) systems, where projected entangled pair states~\cite{verstraete2004renormalization,niggemann1997quantum,nishino1998density,sierra1998density} (PEPSs) are the efficient representation of certain area law states.

Given such variational states, algorithms to find the ground state within such ansatz are of great importance for practical purposes. 
The famous density matrix renormalization group (DMRG) method is an example of an efficient variational algorithm based on the MPS ansatz for 1D systems~\cite{white1992density}. 
For TNSs in 2D, difficulties stem from the fact that the exact contraction of a general TNS, e.g.,computing the norm, scales exponentially $\mathcal{O}(D^L)$ in the linear dimension $L$, where $D$ is the bond dimension~\cite{verstraete2004renormalization,schuch2007computational}. 
A way to overcome this is to perform an approximate contraction of the network---which is possible for certain classes of TNSs~\cite{verstraete2004renormalization,lubasch2014algorithms,nishino1996corner,orus2009simulation}.
This has led to the development of various efficient algorithms that allow one to obtain ground states, using for example imaginary time evolution~\cite{phien2015infinite} or variational energy minimization~\cite{vanderstraeten2016gradient,corboz2016variational,liao2019differentiable}.
While the scaling with bond dimension is not exponential, it is still expensive as it scales with high powers, e.g., $\mathcal{O}(D^{10})$ for a full update in imaginary time evolution \cite{lubasch2014algorithms} and $\mathcal{O}(D^{12})$ for variational energy minimization~\cite{verstraete2004renormalization}.
Moreover, a DMRG-like variational energy minimization algorithm often suffers from being ill-conditioned, which requires an \textit{ad hoc} gauge fixing to solve the problem.
A promising alternative approach combines variational Monte Carlo (VMC) with single layer contraction~\cite{liu2017gradient,liu2019accurate}.
The computational complexity of the algorithm is $\mathcal{O}(N_{MC}D^6)$.
However, the number of samples $N_{MC}$ required in the end of the optimization could also be high, $\sim$10$^5$, for obtaining sufficient accurate gradients.
Recently, it is shown that this can be improved by the direct sampling approach~\cite{vieijra2021direct}.
It remains an important task developing alternative approaches for finding ground state of 2D systems.

In addition to ground state properties, dynamical responses and non-equilibrium physics can also be studied with the help of TNSs.
Simulating real-time evolution is generically difficult because of the fast entanglement growth~\cite{calabrese2005evolution,kim2013ballistic}. 
Nevertheless, TNSs approaches can give reasonable results for certain 2D systems~\cite{murg2007variational,paeckel2019time,PhysRevB.102.035115}.
For example, infinite TNSs with the corner-transfer-matrix renormalization group approach were used to study quantum quench dynamics of the 2D transverse field Ising model~\cite{czarnik2018time,czarnik2019time,dziarmaga2022time} and the hole motion in the $t-J$ model~\cite{10.21468/SciPostPhys.8.2.021}.
Simulating real-time evolution also provides information about the low-lying spectrum through the computation of, e.g., the dynamical spin structure factor.
It is also worth noting that it is possible to directly access the excitation spectrum without simulating the dynamics by variational principle with translational invariant ansatz constructed from iPEPSs~\cite{vanderstraeten2019simulating}.

Recently isometric tensor network states (isoTNSs), which are a subclass of general TNSs, have been proposed as a promising ansatz for efficient algorithms~\cite{zaletel2020isometric,haghshenas2019conversion,hyatt2019dmrg}.
The class of isoTNSs generalizes the isometric condition of the canonical form of MPSs to higher-dimensional tensor networks. 
Using this ansatz, certain computations are significantly faster than in unconstrained TNSs, e.g, the cost of the full-update is reduced from $\mathcal{O}(D^{10})$ to $\mathcal{O}(D^7)$.
However, as isoTNSs are a restricted subclass of TNSs, it is still unclear what quantum phases they can accurately represent.
Recent work~\cite{soejima2020isometric} has shown that all states obtained by applying a finite depth circuit to  a string-net liquid admit exact isoTNS representations. 
In this work, we implement the time evolving block decimation algorithm with isoTNSs ($\text{TEBD}^2$) and introduce a generalized density matrix renormalization group algorithm based on isoTNSs ($\text{DMRG}^2$).
We show that one can study ground state properties and simulate the dynamics of the 2D systems with isoTNSs.

The paper is organized as follows. 
First, we review the basics of isoTNSs and their general properties.
Then we introduce the elementary steps in isoTNSs-based algorithms in Sec.~\ref{sec:isoTNS}.
In this context, we discuss technical details of the ``Moses move'' (MM)~\cite{zaletel2020isometric} and different disentangling approaches--which are essential for efficient handling of isoTNSs. 
Then, we introduce isoTNSs algorithms in Sec.~\ref{sec:Algorithm}.
This includes the $\text{TEBD}^2$ for real and imaginary time evolution and the $\text{DMRG}^2$ for variational ground states search of 2D lattice systems.
In Sec.~\ref{sec:Result}, we demonstrate both methods introduced by computing the dynamical spin structure factor of 2D quantum spin systems.
First, we compare our results for the transverse field Ising model on a square lattice with the prediction by spin-wave theory. 
Second, we consider the Kitaev model on the honeycomb lattice and compare it to the result from the exact solution.
We conclude with a discussion of these results in Sec.~\ref{sec: Conclusion}.

\section{\label{sec:isoTNS} Isometric Tensor Network States}
A pure state describing a quantum many-body system is represented as
\begin{equation}
    \ket{\psi}= \sum_{\sigma_1,\sigma_2 \cdots ,\sigma_N} \Psi^{\sigma_1 \sigma_2 \cdots \sigma_N} \ket{ \sigma_1} \otimes \ket{ \sigma_2} \otimes \cdots \otimes \ket{\sigma_N} ,
\end{equation}
where $\{ \ket{\sigma_i} \in \mathcal{H}_i\}$ are the local basis states.
The full Hilbert space is the tensor product of a set of local Hilbert spaces $\mathcal{H}=\otimes_{i}^N \mathcal{H}_i$. 
The order-N coefficient tensor $\Psi^{\sigma_1 \sigma_2 \cdots \sigma_N}$ contains the full information of the state and the number of parameters scales exponentially $\mathcal{O}(\prod_{i=1}^{i=N} d_i)$ with the system size $N$.
To overcome the exponential scaling, tensor network methods approximate the full tensor $\Psi^{\sigma_1 \sigma_2 \cdots \sigma_N}$ by low-rank tensor decompositions~\cite{hackbusch2012tensor}. 
Usually tensor networks, for example, MPSs in 1D and TNSs in 2D, have a connectivity resembling the underlying lattices and are generically efficient to represent area law states with $\text{poly}(N)$ number of parameters.
In 1D, all states with area law entanglement can be expressed as MPSs with a system size independent bond dimension~\cite{Schuch08}.
In contrast for $D\geq2$, TNSs capture only part of the area law states~\mbox{\cite{ge2016area}}. 
For example, it is still an open question to which extent non-critical chiral topological states can be represented by TNSs in the 2D thermodynamic limit~\mbox{\cite{PhysRevB.92.205307,PhysRevB.95.115309,PhysRevLett.111.236805,PhysRevLett.114.106803,PhysRevB.91.224431}}.

\begin{figure}[t]
\includegraphics[width=0.48\textwidth]{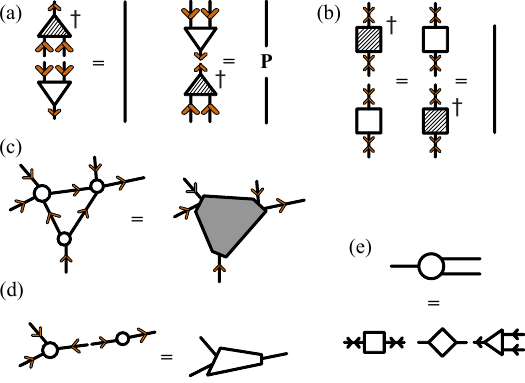}%
\caption{\label{fig:isoTN} (a) The contraction of all incoming (outgoing) arrows of an isometric tensor with its complex conjugate gives an identity (projection) operator. (b) For unitary, both contraction gives an identity operator. Notice that the direction of the arrow is not changed by complex conjugation. (c) From the definition, one can combine isometric tensors of consistent isometric directions to form a larger isometric tensor. (d) When the direction of the arrows does not match, the contraction will return a larger general tensor. (e) SVD replace a general tensor with the contraction of a unitary, a diagonal matrix, and an isometry.}
\end{figure}

Before discussing the details of isoTNSs, we start by giving a brief definition of isometries and introduce the conventions used.
An isometry is a linear map $W: V_s\rightarrow V_l$ from a smaller vector space $V_s$ to a larger vector space $V_l$, such that $W^\dagger W = \mathbb{1},\  W W^\dagger = \mathcal{P}_{V_s}$, where $\mathcal{P}_{V_s}$ is the projection operator to the vector space $V_s$. 
Isometric tensors are tensors that by grouping the legs, i.e. matricization, become isometries.
We consider a convention as shown in Fig.~\ref{fig:isoTN}. 
In particular, for isometric tensors, we draw the indices belonging to the larger dimensions as incoming arrows and the indices for smaller dimensions as outgoing arrows. 
For the unitary matrix, the indices would all be bi-directional. 
For general (non isometric) tensors, we draw the indices without arrows. 
Roughly speaking, isometric tensor networks are then tensor networks consisting of isometric tensors whose edges can consistently be assigned arrows.

\subsection{IsoTNSs in 1D}

We first review the basics of MPSs and identify the MPSs with left/right-normalized tensors as 1D isoTNSs.
For more detailed reviews for MPSs and general TNSs, we refer the readers to~\cite{schollwock2011density,bridgeman2017hand,hauschild2018efficient,paeckel2019time}.

An MPS is defined as
\begin{equation}
    \ket{\psi}= \sum_{\bm \sigma}\sum_{m_1, m_2,\ldots } T^{\sigma_1}_{m_0,m_1}T^{\sigma_2}_{m_1,m_2} \cdots T^{\sigma_N}_{m_{N-1},m_N} \ket{\bm \sigma} , 
\end{equation}
where the $T$ are general tensors of order three and the $m_i$ are indices in virtual space between site $i$ and $i+1$.
The bond dimension $D$ is the dimension of the virtual space, such that $m_i = 1,\cdots ,D$.
Carrying out the summation $\sum_{m_1, m_2,\cdots }$, i.e. contraction, over the virtual indices explicitly gives the coefficient $\Psi^{\sigma_1 \sigma_2 \cdots \sigma_N}$. 
The number of parameters of MPSs scales as $\mathcal{O}(ND^2)$, which avoids the exponential scaling in the system size.

One can always transform the MPS tensors into the left/right-normalized form without loss of generality.
Importantly, the left/right normalization of the tensors simplifies and stabilizes MPS based algorithms.
A tensor $A^{\sigma_i}_{m_{i-1},m_{i}}$ is left-normalized if it satisfies the isometric condition,
\begin{equation}\label{MPS_left}
    \sum_{\sigma_i m_{i-1} } A^{\sigma_i}_{m_{i-1},m_{i}}  \left ( A^{\sigma_i}_{m_{i-1},m'_{i}} \right)^* = \mathbb{1}_{m_{i},m'_{i}} .
\end{equation}
Similarly a tensor $B^{\sigma_i}_{m_{i-1},m_{i}}$ is right-normalized if it satisfies the isometric condition, 
\begin{equation}\label{MPS_right}
    \sum_{\sigma_i m_{i} } B^{\sigma_i}_{m_{i-1},m_{i}} \left ( B^{\sigma_i}_{m'_{i-1},m_{i}} \right )^*  = \mathbb{1}_{m_{i-1},m'_{i-1}} .
\end{equation}

The tools to bring MPSs tensors to the left/right-normalized form are orthogonal matrix decompositions, i.e. QR and SVD. 
Given an MPS with general tensors, we can bring all tensors successively into the left/right-normalized form by successive SVDs or QR decompositions.
For example, one can start from the left with QR decomposition $T^{\sigma_1}_{m_0,m_1} = \sum_{m_{1'}} A^{\sigma_1}_{m_0,m_{1'}} R_{m_{1'},m_{1}}$, and combine $\sum_{m_{1}} R_{m_{1'},m_{1}} T^{\sigma_2}_{m_1,m_2}= \tilde{T}^{\sigma_2}_{m_{1'},m_2}$.
Now the original $T^{\sigma_1}_{m_0,m_1}$ tensor becomes a left-normalized tensor $A^{\sigma_1}_{m_0,m_{1'}}$.
Iteratively, one can \emph{exactly} bring all tensors into the left-normalized form.
Similarly, one can start from the right and move left and end up with all tensors being right-normalized.

The orthogonality center of an MPS is a single bond or a region of sites such that to the left of the center all tensors are left-normalized and to the right all are right-normalized (note that this does not have to be the geometric center of the chain). 
By the combination of both moves mentioned above, one can \emph{exactly} obtain an MPS with normalized tensors and orthogonality center at any desired bond or region.

For example, following the normalization procedure by QR decomposition mentioned above from both ends of the MPS inwards, we have the following decomposition:
\begin{equation}
\begin{split}
\label{MPS_mixed_bond}
    &\Psi^{\sigma_1 \cdots \sigma_N} \\
    &\ \ = \sum_{\{m_i\} } A^{\sigma_1}_{m_1}\cdots A^{\sigma_{l-1}}_{m_{l-1},m_{l}}  R_{m_l,m_{l'}} B^{\sigma_{l}}_{m_{l'},m_{l+1}}\cdots B^{\sigma_N}_{m_{N-1}}\\
\end{split} .
\end{equation}
By definition, $R_{m_l,m_{l'}}$ is the orthogonality center on bond-$l$.
Note that, the orthogonality center on bond-$l$ can be a general matrix without the restriction of upper-triangular form.
From this point on, we denote such general matrix by $\Psi_{m_l,m_{l'}}$, which is also known as the 0-site wavefunction.

We can obtain the MPS with orthogonality center on a single site by merging %
$\sum_{m_{l'}} \Psi_{m_l, m_{l'}} B^{\sigma_{l}}_{m_{l'},m_{l+1}} = \Psi^{\sigma_{l}}_{m_{l},m_{l+1}}$,
leading to
\begin{equation}\label{MPS_mixed_site}
\begin{split}
&\Psi^{\sigma_1 \cdots \sigma_N}   =\\
&\ \ \sum_{\{m_i\} } A^{\sigma_1}_{m_1}\cdots A^{\sigma_{l-1}}_{m_{l-1},m_{l}} \Psi^{\sigma_{l}}_{m_{l},m_{l+1}} B^{\sigma_{l+1}}_{m_{l+1},m_{l+2}} \cdots B^{\sigma_N}_{m_{N-1}} .
\end{split}
\end{equation}
$\Psi^{\sigma_{l}}_{m_{l},m_{l+1}} $ is the orthogonality center on site-$l$ and a single-site wavefunction.
We can move the orthogonality center forward keeping constant bond dimension by repeating orthogonal matrix decompositions and merging tensors.

The tensors excluding the orthogonality center, for example $\Psi^{\sigma_{l}}_{m_{l},m_{l+1}}$ in  Eq.~\eqref{MPS_mixed_site}, are a collection of isometries
\begin{equation*}
    \{ A^{\sigma_1}_{m_1}, \cdots,  A^{\sigma_{l-1}}_{m_{l-1},m_{l}},B^{\sigma_{l+1}}_{m_{l+1},m_{l+2}}, \cdots, B^{\sigma_N}_{m_{N-1}} \}.
\end{equation*}
Contracting all the internal virtual indices, they form a single isometry $T^{V \leftarrow \partial V}$.
In other words, the boundary map $T^{V \leftarrow \partial V}$ of the orthogonality center $\Psi^{\sigma_{l}}_{m_{l},m_{l+1}}$ is an isometry mapping from the virtual space $\partial V$ to physical Hilbert space $V$.
Similarly, the boundary map of the orthogonality center on bond-$l$ is also isometric.
Because the boundary map is isometric, we can interpret the orthogonality center itself as the wavefunction in the lower dimensional space.
For example, the $\Psi_{m_l,m_{l'}}$ can be interpreted as a 0D representation of the state in terms of orthogonal states $|m_l\rangle$ and $|m_{l'}\rangle$, i.e.,
\begin{align} \label{MPS-0D-WF}
\ket{\psi} &= \sum_{{\bm \sigma}}\sum_{\{m_i\} } A^{\sigma_1}_{m_1}\cdots A^{\sigma_{l-1}}_{m_{l-1},m_{l}}\ket{\sigma_1, \ldots, \sigma_{l-1}} \nonumber \\
&\qquad \times   \Psi_{m_l,m_{l'}} B^{\sigma_{l}}_{m_{l'},m_{l+1}}\cdots B^{\sigma_N}_{m_{N-1}}  \ket{\sigma_l, \ldots, \sigma_{N}} \nonumber \\
&= \sum_{m_l,m_{l'}} \Psi_{m_l,m_{l'}} |m_l\rangle |m_{l'}\rangle.
\end{align}
Since the isometric map is norm-preserving, the truncation on $\Psi_{m_l,m_{l'}}$ based on an SVD is not only optimal for $\Psi_{m_l,m_{l'}}$ in $\mathcal{L}_2$ norm but also for $\Psi^{\sigma_1 \cdots \sigma_N}$ in $\mathcal{L}_2$ norm.

Moreover, we define the norm tensor $N_l$ with respect to site-$l$ as the contraction of the norm $\braket{\Psi|\Psi}$ but leaving out tensors on site-$l$, e.g., $\Psi^{\sigma_{l}}_{m_{l},m_{l+1}}$ and $ ( \Psi^{\sigma_{l}}_{m_{l},m_{l+1}}  )^*$.
In other words, it is the contraction of the boundary map $(T^{V \leftarrow \partial V})^\dagger T^{V \leftarrow \partial V}$.
Because of the isometric condition of the boundary map, the norm tensor $N_l =   (T^{V \leftarrow \partial V})^\dagger T^{V \leftarrow \partial V} = \mathbb{1}_{\partial V}$ with respect to the orthogonality center is an identity operator.

\begin{figure*}[t]
\includegraphics[width=0.95\textwidth]{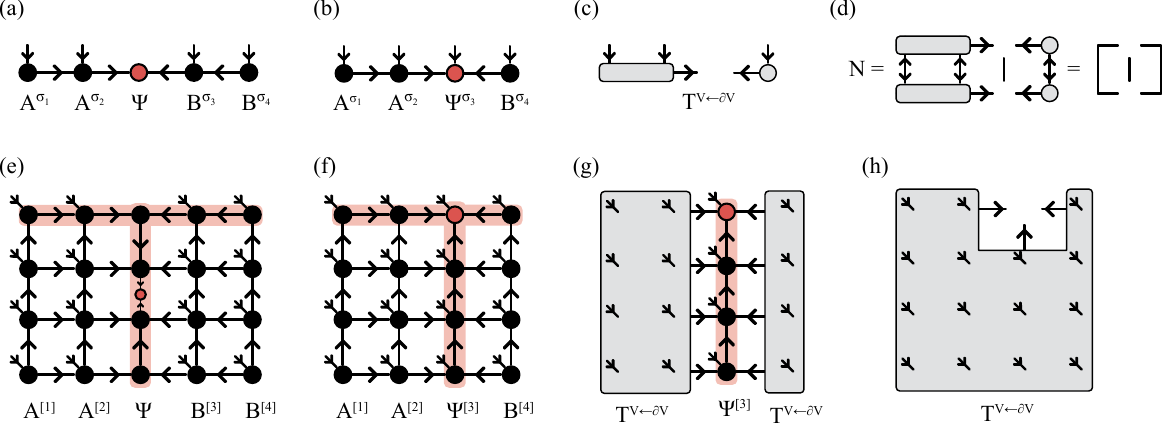}
\caption{\label{fig:isoTN2}
(a) An MPS with orthogonality center on a bond  and (b) on a single-site .
(c) The boundary map $T^{V\leftarrow \partial V}$ of the MPS (b).
(d) The norm tensor $N$ of the orthogonality center $\Psi$ of the MPS (b).
Analogously, an isoTNS in 2D having orthogonality hypersurface colored in red and with a column (e) without physical indices $\Psi$ and (f) with physical indices $\Psi^{[l]}$.
(g) The decomposition of an isoTNS as the left and right isometric boundary maps with effective 1D wavefunction $\Psi^{[3]}$.
(h) The boundary map $T^{V\leftarrow \partial V}$ of the orthogonality center of the isoTNS (f).
}
\end{figure*}

From the above we see that an isoTNS in 1D has the following properties:
It is a tensor network composed of isometries, as from Eq.~\eqref{MPS_left} and~\eqref{MPS_right}, and all the tensors excluding the orthogonality center form an isometric boundary maps in both Eq.~\eqref{MPS_mixed_bond} and ~\eqref{MPS_mixed_site}.

In passing we mention that the canonical form of MPSs imposes an additional condition aside from the isometric condition~\cite{perez2007matrix},
\begin{equation}\label{MPS_left_canonical}
    \sum_{\sigma_i m_i }  A^{\sigma_i}_{m_{i-1},m_{i}} \rho^{(i)}_{m_i,m_i} \left ( A^{\sigma_i}_{m'_{i-1},m_{i}} \right)^* = \rho^{(i-1)}_{m_{i-1},m'_{i-1}}
\end{equation}
where the $\rho^{(i)}$ and $\rho^{(i-1)}$ are positive diagonal matrices and $\text{Tr}\left[ \rho^{(i)} \right]=\text{Tr}\left[ \rho^{(i-1)} \right]=1$.
An MPS with all left/right-normalized tensors can be brought into canonical form by the unitary transform determined by the condition in Eq.~\eqref{MPS_left_canonical}.
It is often stated in the literature that the canonical form of MPSs is crucial to the success of the DMRG and TEBD algorithms.
Here, we would like to point out most of the advantages in 1D numerical algorithms come from merely the isometric condition.
In terms of numerical algorithms, the linear geometry of MPSs itself ensures the exact shifting of orthogonality center and the exact contraction between MPSs with cost $\mathcal{O}(D^3)$.
\rewrite{}{Apart from that, the success of efficient 1D algorithms including DMRG and TEBD with MPSs are because of the two properties at the orthogonality center: (i) the identity norm tensor $N$ and (ii) the optimal bond truncation.
Firstly, because the norm matrix at the orthogonality center is the identity, the optimization problem in DMRG is a standard eigenvalue problem instead of a generalized eigenvalue problem. 
Secondly, at each step in the algorithm, the truncation based on the SVD over the two-site orthogonality center is a local update, which is optimal for the global state since the basis is orthonormal.
Both properties above come from the isometric condition of the normalized MPS tensors as discussed above.}
And it does not require the additional Schmidt-state gauge condition from the canonical form.
To generalize the success of 1D algorithms to higher dimensions, we focus on keeping the isometric condition of the TNSs.

\subsection{IsoTNSs in higher dimensions \label{subsec: isoTNS_2}}

To generalize the above framework to higher dimensions, we consider isoTNSs with an arrangement of isometries described as follows.
For a $k$-dimensional isoTNS, we assume that we can find a $(k-1)$-dimensional hyperplane in the isoTNS, that separates the $k$-dimensional space into two parts with the condition that the isometries from both parts are pointing towards the hyperplane.
This means we can define a $(k-1)$-dimensional state in terms of the boundary states of the two isometric maps representing the full $k$-dimensional state.
We can then successive continue this reduction of the dimension until 0D.
For clarity, let us first consider some examples:
(i) A 1D MPS with an orthogonality center, as discussed above, has a 0D orthogonality center and 1D boundary maps. The truncation on the orthogonality center is optimal for the 1D quantum state as shown in Eq.~\eqref{MPS-0D-WF}. (ii) A 2D isoTNS on a rectangular lattice as shown in Fig.~\ref{fig:isoTN2}e. The column colored in red is an effective 1D wavefunction with 2D isometric boundary maps as shown in Fig.~\ref{fig:isoTN2}g.
We can view the 1D wavefunction as an 1D MPS by grouping and reinterpreting the virtual indices.

With the structure of isoTNSs in mind, we now discuss the general properties of isoTNSs and give concrete examples using 2D isoTNSs.
In terms of the numerical algorithm, isoTNSs have the two ideal properties: (i) \emph{optimal bond truncation} on the orthogonality center and (ii) \emph{identity norm tensor} on the orthogonality center.
These two properties are the direct consequence of the isometric boundary map.
The nested isometric boundary maps of dimension $k, k-1, \ldots, 1$ would form a single isometric boundary maps to the orthogonality center.
Therefore, any optimal truncation on the orthogonality center in the $\mathcal{L}_2$ norm, e.g. SVD, is an optimal truncation of the $k$-dimensional wavefunction because the isometric boundary map is norm-preserving.
Similarly, because of the boundary map of the tensor on the orthogonality center is isometric as shown in Fig.~\ref{fig:isoTN2}h, the norm tensor $N_l =   (T^{V \leftarrow \partial V})^\dagger T^{V \leftarrow \partial V} = \mathbb{1}_{\partial V}$ is an identity operator.
Formally, the arrangement of isometries described above defines a causal structure of the tensor network flowing in the reverse direction of the arrows since the isometries would form Kraus operators. 
With our setup, the arrows of isometries don't form loop. It means the isoTNSs considered are physical states that could be prepared using quantum circuits~\cite{slattery2021,PhysRevLett.128.010607}.

In this work, we focus on 2D isoTNSs and use a notation inherited from MPSs. 
We can view the 2D isoTNS on rectangular lattice as an generalization of the MPS where each isometry of the MPS is extended to a column of isometries.
The 2D isoTNSs have a similar pattern in terms of the isometries as 1D MPSs.
Therefore, we use $A^{[l]}$ and $B^{[l]}$ to denote columns of isometries that are left-normalized and right-normalized. We use $\Psi$ and $\Psi^{[l]}$ to denote the columns containing orthogonality center that are without and with physical indices as shown in Fig.~\ref{fig:isoTN2}e and Fig.~\ref{fig:isoTN2}f.
We see in Fig.~\ref{fig:isoTN2}, that the direction of the isometries is chosen consistently pointing toward the orthogonality center (red circle) and each column can be contracted to an isometry, recovering the MPS structure.

Following the discussion above, we see the red colored region in Fig.~\ref{fig:isoTN2}e and Fig.~\ref{fig:isoTN2}f define sub-regions of isoTNSs which have special properties.
We call these effectively 1D regions the ``orthogonality hypersurfaces'' of the isoTNSs.
Inside this region, we can move the orthogonality center from site to site exactly by orthogonal matrix decomposition.
This 1D region has only incoming arrows and hence an isometric boundary map.
Any variationally optimal algorithm inside the orthogonality hypersurface is variationally optimal for the global state.
Utilizing these properties, we can run similar 1D algorithms on the orthogonality hypersurface.

In the following, we consider algorithms for 2D isoTNSs and work with columns for conceptual convenience.
The column $\Psi^{[l]}$ contains the physical indices and is in the orthogonality hypersurface.
All columns to the left of the column $\Psi^{[l]}$ are the left-normalized columns $A^{[m]}$ and to the right the right-normalized columns $B^{[n]}$.
We have discussed that we can run effectively 1D algorithms on the column $\Psi^{[l]}$ and move the center site, i.e. orthogonality center, freely within the column. 
However, it is not possible to perform  a QR decomposition to the entire column $\Psi^{[l]}$ because of the exponential scaling.
In particular, we cannot directly shift the isometry direction of the whole column $\Psi^{[l]}$ to $A^{[l]}\Psi$ by a simple orthogonal matrix decomposition and thus we require some new algorithms.
We describe two different ways to do this in \mbox{Sec.~\ref{subsec:shift_column_var} and Sec.~\ref{subsec: MM}.}

General 2D TNSs contractions have exponential complexity and require environment approximations~\cite{verstraete2004renormalization} using for example boundary MPSs or corner transfer matrices.
For a 2D isoTNS, the expectation values for any operator acting only on column $\Psi^{[l]}$ permits an exact evaluation with polynomial scaling in system size. 
The expectation values can be evaluated using standard MPS contraction on this column, i.e. $\braket{\psi | \hat{O} | \psi} = \braket{\Psi^{[l]} | \hat{O} | \Psi^{[l]}} $---no contraction of 2D networks is involved. 
For the evaluation of observable outside the column $\Psi^{[l]}$, one can shift the orthogonality hypersurface (see Sec.~\ref{subsec:shift_column_var}) and evaluate the expectation as described above.

\subsection{\label{subsec:shift_column_var} Variational Moses move}

\begin{figure}[b]
\includegraphics[width=0.99\columnwidth]{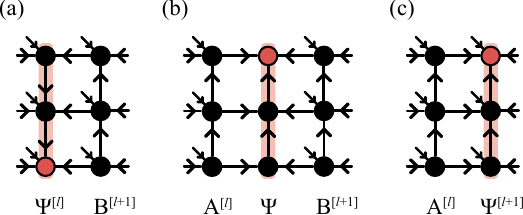}
\caption{\label{fig:MM_move} The column $\Psi^{[l]}$ is shifted by starting with a Moses move from (a) $\Psi^{[l]}B^{[l+1]}$ to (b) $A^{[l]} \Psi B^{[l+1]}$. The combination of $\Psi B^{[l+1]} = \Psi^{[l+1]}$ would lead to higher bond dimension which requires further truncation. We implement this with the same idea as the standard MPO-MPS compression method. This gives the two columns $A^{[l]}\Psi^{[l+1]}$ in (c). An optional but more expensive step includes the maximizing of the two columns overlap $\braket{\psi|\psi'}$ of the wavefunction before in (a) and after in (c). We describe the detail of the steps above in Appendix~\ref{appendix:detail_MM}. }
\end{figure}

To implement practical algorithms using isoTNSs, we need a method to move the column $\Psi^{[l]}$ around efficiently. 
Recall in 1D the orthogonal matrix decompositions, e.g. QR decomposition, move the orthogonality center (thus shift the directions of the isometries) by $\Psi^{\sigma_{l-1}} B^{\sigma_{l}} = A^{\sigma_{l-1}} \Psi B^{\sigma_{l}} = A^{\sigma_{l-1}} \Psi^{\sigma_{l}}$, where each of the $\Psi, A, B$ is a tensor.
In 2D, $\Psi, A, B$ each refers to a column of tensors as in Fig.~\ref{fig:isoTN2}. 
There is no orthogonal decomposition of the whole column of tensors while keeping the locality (tensor decomposition) structure. 
QR decomposition or SVD on individual tensor would destroy the matching of isometry directions and thus destroy the identity boundary map of isoTNS.

While an exact solution does not exist, we consider the following variational problem:
Given the column $\Psi^{[l]}$, we find columns $A^{[l]}, \Psi$ such that the distance between two represented states is minimized, i.e.,
\begin{equation}\label{variational_MM}
    \argmin_{A^{[l]}, \Psi}  \left \lVert \Psi^{[l]} - A^{[l]} \Psi   \right \rVert^2
\end{equation}
with the constraint that $A^{[l]}$ is a column of isometries pointing toward $\Psi$ column (see Fig.~\ref{fig:MM_move}). 
Notice that the $\Psi$ column does not have physical indices.
One can think of this variational problem as the analogy of the QR decomposition for a column of tensors except that $\Psi$ is not restricted to be upper triangular. 
While the QR decomposition is a deterministic algorithm providing a numerical exact decomposition, the variational problem here generically only provides an approximate decomposition. 
If the problem could be solved, the next step is to contract the ``zero-site" column $\Psi$ to the next right-normalized column $B^{[l+1]}$ and form the $\Psi^{[l+1]}$ column as in Fig.~\ref{fig:MM_move}b and Fig.~\ref{fig:MM_move}c. 
And one can continue forward with the move in analogy to the 1D case. We call this procedure of solving Eq.~\eqref{variational_MM} iteratively and obtaining $A^{[l]}, \Psi$ columns the \emph{variational Moses move}.

We separate the variational Moses move into two parts which correspond to two common types of variational problems for tensor networks. 
The first part is the variational optimization over tensors in column $\Psi$. 
This is an unconstrained optimization problem with general tensors. 
The second part is the variational optimization over the isometries in column $A^{[l]}$. 
This is a constrained optimization problem with isometry tensors. 
Both type of problems could be solved in an alternating least-squares fashion as described below.

\subsubsection{ Unconstrained optimization for tensors in column $\Psi$}

The unconstrained optimization problem over $\Psi$ is similar to the problem of variational approximation that occurs in the iterative compression of an MPS~\cite{schollwock2011density,lubasch2014algorithms}. 
The general problem has the following setup:
Given a target state $\ket{\psi}$, we want to find the optimal isoTNS $\ket{\phi}$ representation in $\mathcal{L}_2$ norm by varying a single tensor $x$ at site-$l$ at a time 
\footnote{Notice that when the problem is imposed on normalized tensor networks, minimizing the difference in $\mathcal{L}_2$ norm is equivalent to maximizing the fidelity $\mathcal{F}(\phi, \psi) = \lvert \braket{\phi | \psi} \rvert^2$. This is because of the global phase is not fixed and one can always set it such that $ \lvert \braket{\phi | \psi} \rvert= \mathrm{Re}\left[ \braket{\phi | \psi} \right]$}.

The solution of the minimization problem
\begin{equation}\label{eq:distance_two_state}
    \argmin_{x} \Big\lVert |\psi\rangle - |\phi(x)\rangle  \Big\rVert ^2
\end{equation}
must satisfy the extremum condition $
\partial_{x^*}  \braket{\phi|\phi} - 
\partial_{x^*}  \braket{\phi|\psi} = 0 $.
Therefore, the optimal tensor $x$ is found by solving the system of linear equations,
\begin{equation}\label{eq:sol_distance}
    N_{l}x = b,\ \ \mathrm{i.e.}\  x = N_{l}^{-1}b
\end{equation}
where $N_{l}$ is the  norm matrix obtained from contraction $\braket{\phi|\phi}$ and leaving out tensor $x$ in ket and tensor $x^*$ in bra, and $b$ results from leaving out tensor ${x^*}$ from $\braket{\phi|\psi}$.

Similar to the variational compression of MPS, at each update, we keep the update tensor $x$ at the orthogonality center such that the norm matrix is the identity operator, $N_{l} = \mathbb{1}$.
As a result, the optimal update is given by the contraction $b$ without solving a system of linear equations.
In variational Moses move, each local update $b$ is formed by the contraction of three column $\Psi^{[l]}, A^{[l]}, \Psi$.
After each update, we move the orthogonality center to the next site in the column by orthogonal matrix decomposition.

\subsubsection{Constrained optimization for isometries in column $A^{[l]}$}

The constrained optimization over tensors $x$ within column $A^{[l]}$ requires $x$ to be an isometry. 
This could not be solved by the same approach as in Eq.~\eqref{eq:sol_distance}, since the solution is a general tensor. 
The isometric constraint cannot be consistently restored by orthogonal matrix decomposition.
First, we rewrite the problem from minimizing $\mathcal{L}_2$ distance between states to maximizing the real part of the overlap,
\begin{align*}
&\argmin_{x \in \mathrm{isometry}}\    2 - 2 \mathrm{Re} \left [ \braket{\psi | \phi} \right ] \\
= &\argmax_{x \in \mathrm{isometry}}\   \mathrm{Re} \left [ \braket{\psi | \phi} \right ] \\
= &\argmax_{x \in \mathrm{isometry}}\    \mathrm{Re} \left [ Tr[b^\dagger x] \right ]. 
\end{align*}
In the first line, we use the condition that the isoTNSs have identity norm. 
In the last line, we reshape the isometric tensor $x$ and the tensor $b$ to matrices.
The resulting constrained optimization problem is known as the orthogonal Procrustes~\cite{gower2004procrustes}~\footnote{The problem is named after Procrustes, a bandit from the Greek mythology, who forced passersby to fit to his bed by stretching or cutting off their bodies.} problem and permits closed form solution.
The optimal update for $x$ is given by $VU^\dagger$ from the SVD of the $b^\dagger=U\Sigma V^\dagger$ matrix.
A similar problem appears also in the optimization of multi-scale entanglement renormalization ansatz (MERA) with linearization~\cite{evenbly2009algorithms,evenbly2014algorithms}. 
The derivation and detailed discussion are given in Appendix~\ref{appendix:opt_iso}.

One could start with randomly initialized $A^{[l]}, \Psi$ and iteratively sweep through and update all the tensors in both columns.
The algorithm stops when desired accuracy or convergence criterion is reached.
In practice, we observe that random initialization with local updates may lead to slow convergence toward a sub-optimal minimum. 
Therefore, we introduce in the next section a complementary approach for shifting the column, which could serve as a good initialization for variational Moses move.

After each variational MM, one contracts the $\Psi$ column to the next column. 
The two columns contraction is similar to the application of an MPO to an MPS.
Therefore, the most efficient way is to consider the MPO-MPS contraction variationally~\cite{stoudenmire2010minimally,paeckel2019time}.
Note that there are a few subtleties:
The variational Moses move $\Psi^{[l-1]} B^{[l]} = A^{[l]} \Psi^{[l]}$ is generically not exact and inherits errors. 
Furthermore, after the move and tagging the column, the bond dimension on column $l$ grows. 
To keep the bond dimension fixed, a truncation occurs.

\subsection{Sequential Moses move \label{subsec: MM}}

In this section, we review a sequential solution for moving the orthogonality hypersurface~\cite{zaletel2020isometric}. 
The sequential MM is a greedy algorithm that sequentially splits one column into two satisfying the isometric constraints by a single unzipping sweep. 
We observe in practice that the sequential MM has an error very close to the optimal variational result while being much faster.
In addition, this approximate solution can serve as a good initialization for variational Moses move.

\begin{figure}[t]
\includegraphics[width=1.\columnwidth]{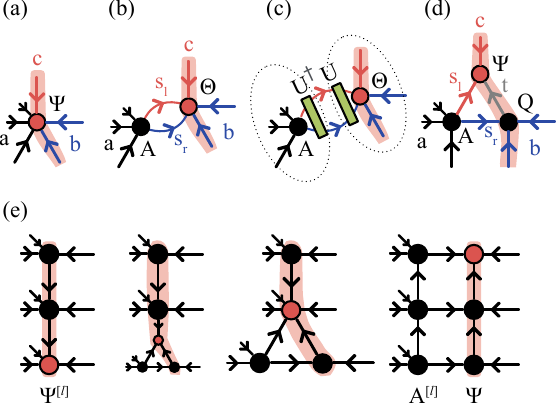}
\caption{\label{fig:splitter} (a-d) The tripartite decomposition (e) Moses move $\Psi^{[l]} = A^{[l]} \Psi$ by iterative tripartite decomposition.
(a) We group the indices of color black, blue, red together respectively as $a,b,c$ and denote the orthogonality center as $\Psi_{a,b,c}$.
(b) After the SVD and merging the singular values to the orthogonality center, we again group the indices of same color together and denote it as $\Theta_{(s_l,c), (s_r,b)}$. The index $s_r$ is colored in blue and $s_l$ in red.
(c) The insertion of identity operator $I=U^\dagger U$, where $U$ is chosen to minimize the entanglement in $\Theta$. We absorb the $U^\dagger$ into isometry $A$ and $U$ into $\Theta$.
(d) SVD on the final $\Theta$ complete the tripartite decomposition.
(e) Utilize the decomposition, we unzip the columns with orthogonality center moving to the top.
}
\end{figure}

The idea of the sequential MM is to perform a sequence of tripartite decompositions at the orthogonality center.
We illustrate the idea in Fig.~\ref{fig:splitter}.
At each step, we split the single tensor into three tensors as shown in Fig.~\ref{fig:splitter}a to Fig.~\ref{fig:splitter}d.
Iteratively, we split the full column into two by repeat such decompositions as illustrated in Fig.~\ref{fig:splitter}e.
To simplify the notation, we always merge the indices of the tensor at the orthogonality center as an order-3 tensor and denote it as $\Psi_{a,b,c}$.

The tripartite decomposition is composed of two consecutive SVDs and a gauge fixing procedure.
We use Einstein summation convention and describe the decomposition step by step as follows:
\begin{enumerate}[label=(\roman*)]
    \item Perform an SVD on $\Psi_{(a),(b,c)}$
    \begin{equation*}
        \Psi_{a,(b,c)} = A_{a,s} \Lambda_{s,s} V_{s,(b,c)} = A_{a,s} \Theta_{s,b,c}.
    \end{equation*}
    The $A$ is an isometry.
    In principle, we could also directly arrive at the final form by reduced QR decomposition  $\Psi_{(a),(b,c)}=A_{a,s} \Theta_{s,(b,c)}$.
    However, in certain cases we would truncate the bond dimension which requires the SVD instead.
 
    \item 
    Split the index $s$.
    To get to Fig.~\ref{fig:splitter}b, we split the index $s$ into $s_l, s_r$.
    In practice, we choose their dimensions $|s_l| \sim |s_r| \sim \lfloor \sqrt{|s|} \rfloor$ in order to distribute the bonds evenly (though in anisotropic models, other prescriptions may be appropriate).
    Also the bond dimension $ | s_l |$ and $| s_r |$ should be smaller than the maximum vertical bond dimension $D_\text{V}$ and horizontal bond dimension $D_\text{H}$ set for the simulation.
    Notice that if $|s| > |s_l| |s_r|$, then the SVD in step (i) is a truncated SVD, i.e.
    \begin{equation}\label{first_svd}
    \Psi_{a,b,c}\approx A_{a,(s_l,s_r)} \Theta_{(s_l,s_r),b,c}.
    \end{equation}
    For a given pairs of $(s_l, s_r)$, we first truncate and keep only the leading $|s_l| |s_r|$ singular values in $\Lambda_{s,s}$.
    We then form the $\Theta_{s,b,c}$ and reshape it into $\Theta_{s_l, s_r, b, c}$.
    Note the ordering of the reshaping, column-major or row-major, has only minor effect for the following reason.
    The decomposition $\Psi = A \Theta$ has a gauge degree of freedom $A \Theta= (A U^\dagger) (U \Theta) = A' \Theta'$, where $U$ is an arbitrary unitary matrix.
    Absorbing $U^\dagger$ in to orthonormal basis $A$ results in a new orthonormal basis $A'$.
    We defer the discussion of finding the unitary $U$, i.e. fixing the gauge, for the decomposition until after we describe the full picture of the sequential MM.
    At this point, we assume we fix the gauge by the unitary determined by some procedures.
    
    \item Perform an SVD on $\Theta_{(s_l,c), (s_r,b)}$.
    Given $\Theta_{s_l,s_r,b,c}$, we rearrange and group the indices $(s_l,c)$ and $(s_r, b)$ and perform the (truncated) SVD as shown in Fig.~\ref{fig:splitter}b and Fig.~\ref{fig:splitter}d. That is
    \begin{align}\label{second_svd}
    \Theta_{(s_l,c), (s_r,b)} \simeq V_{(s_l,c),t} \Lambda_{t,t} Q_{t, (s_r,b)} = \Psi_{s_l,c,t} Q_{t, (s_r,b)}.\nonumber\\
    \end{align}
    Finally, combining Eq.~\eqref{first_svd} and Eq.~\eqref{second_svd}, we have the decomposition 
    \begin{equation}
    \Psi_{a,b,c} \approx A_{a,s_l,s_r} \Psi_{s_l,c,t} Q_{t,s_r,b}
    \end{equation}
    as in Fig.~\ref{fig:splitter}d. 
\end{enumerate}

The entire sequential MM is shown in Fig.~\ref{fig:splitter}e, where we start from the orthogonality center at the bottom and move to the top, unzipping the entire $\Psi^{[l]}$ column by the tripartite decomposition. 
Before the MM, the $\Psi^{l]}$ column is composed of tensors $[\Psi^{\sigma_1}, P^{\sigma_2}, \ldots, P^{\sigma_{L_y - 1}}, P^{\sigma_{L_y}} ]$.
At each step, we 
\begin{enumerate}[label=(\roman*)]
    \item Reshape the orthogonality center $\Psi^{\sigma_l}$ to an order-3 tensor $\Psi_{a,b,c}$
    
    \item Perform a tripartite decomposition on $\Psi_{a,b,c}$ as in Fig.~\ref{fig:splitter}a-d
    
    \item Merge the orthogonality center after the decomposition with the tensor above to form the new center, i.e. $\Psi P^{\sigma_{l+1}} = \Psi^{\sigma_{l+1}}$.
\end{enumerate}
After step (iii) (merging the tensors), we recover the form as in Fig.~\ref{fig:splitter}a. Therefore, we can repeat and continue from step (i) until we reach the top of the column.
We end the splitting at the top of the column by a single SVD. 
Collecting the $A$ tensors and the $Q$ tensors along the way, we obtain the $A^{[l]}$ column consisting of tensors $[A^{\sigma_1}, A^{\sigma_2}, \ldots, A^{\sigma_{L_y - 1}}, A^{\sigma_{L_y}} ]$ and the $\Psi$ column consisting of tensors $[Q, Q, \ldots, Q, \Psi]$.
The full procedure is illustrated in Fig.~\ref{fig:splitter}e.
To continue, we see we could combine the $\Psi$ column easily with the $B^{[l+1]}$ column as in Fig.~\ref{fig:MM_move}b and result in the $\Psi^{[l+1]}$ column consisting of tensors $[Q^{\sigma_1}, Q^{\sigma_2}, \ldots, Q^{\sigma_{L_y - 1}}, \Psi^{\sigma_{L_y }}]$ as in Fig.~\ref{fig:MM_move}c.

In practice, we would impose a maximum bond dimension for column $\Psi$ and also for column $\Psi^{[l+1]}=\Psi B^{[l]}$ as in Fig.~\ref{fig:MM_move}c. 
Truncations may take place on column $\Psi$ during the second SVD and on column $\Psi^{[l+1]}$ after the combination of columns.

Now we discuss the criterion and how to find the optimal $U$ in the tripartite decomposition.
The crucial insight from \cite{zaletel2020isometric} is that the truncation error occurs in MM can be made smaller by utilizing the gauge degree of freedom between tensors (see Fig.~\ref{fig:splitter}c).
An insertion of a pair of unitary and its conjugate, i.e. identity operator, before the second SVD leave the overall tensors invariant.
However, this changes the distribution of the singular values of the tensor $\Theta$ when we absorb the unitary $U$ into it and hence changes the truncation error.
Therefore, we include the insertion of the pair of unitary and its conjugate as in Fig.~\ref{fig:splitter}c between the first and second SVD.

The optimal $U$ is the $U$ that leads to the smallest truncation error.
Therefore, we solve for the variational problem
\begin{equation}
    \argmin_{U \in \mathrm{unitary}} \mathcal{L}( U\Theta)
\end{equation}
where the cost function $\mathcal{L}(U\Theta)$ is chosen such that minimizing it reduces truncation error.
In the following, we describe two classes of cost functions.
For all the cost functions considered, the gradient of the unitary
can be computed by analytical derivation or auto-differentiation scheme~\cite{liao2019differentiable}.
One can solve the optimization problem and obtain the optimal unitary with Riemannian gradient descent or Newton-based methods over the Stiefel manifold~\cite{smith1994optimization,abrudan2008steepest,ferris2012variational,hauru2020riemannian,luchnikov2020riemannian}. 
We give an overview of these procedures in Appendix~\ref{appendix:RGD}.

\subsubsection{Entanglement entropies as cost functions}

The first class of cost-functions  we consider are the entanglement entropies.
In~\cite{zaletel2020isometric}, R\'enyi-$\alpha$ entanglement entropies $\frac{\alpha}{1 - \alpha} \log{ [\text{Tr}(\rho^\alpha)] }$ are chosen as the cost function, where $\rho$ is the reduced density matrix from the bipartition of the tensor $\Theta_{(c,s_l),(b,s_r)}$ and $\alpha = \frac{1}{2}$ or $2$.
Similar variational problems of ``disentangling" also show up in various different contexts.
In~\cite{hauschild2018finding}, one finds the minimal entanglement representation of MPSs for purified state by utilizing the gauge degree of freedom of the ancilla space in the purification, where R\'enyi-$2$ entropy is considered.
In a different context of interacting fermionic system, the local mode transformation looks for the optimal unitary leading to smallest truncation in MPS representation by minimizing the R\'enyi-$\frac{1}{2}$ entropy~\cite{krumnow2016fermionic,krumnow2019towards}.

Choosing R\'enyi-$\frac{1}{2}$ entropy as cost function is justified by the fact that R\'enyi-$\alpha$ entropy with $\alpha < 1$ upper bounds the truncation error for a fixed bond dimension~\cite{verstraete2006matrix}.
With $\alpha > 1$, the optimization, however, would not give a certified bound on the truncation error.
Such cost function is still often considered in the literature since the optimization is simpler.
For example, the optimization converges to minimum quickly with the Evenbly-Vidal algorithm~\cite{evenbly2014algorithms} due to the cost function landscape~\cite{luchnikov2020riemannian}. See Appendix~\ref{appendix:G-V}.
Therefore, it is common to consider R\'enyi-$\alpha$ entropy with $\alpha = 2$ as an alternative cost function or an initialization for optimization with $\alpha < 1$.

\subsubsection{Truncation error as cost functions}

The second type of cost functions is simply the truncation error given by the maximum bond dimension $D$
\begin{equation}
    \epsilon_{D} = \sum_{i=D+1}^{\infty} \Lambda_{i,i}^2 .
\end{equation}
Direct minimization of the truncation error is possible and may be more effective than minimizing a upper bound or surrogate cost functions in some cases.

\begin{figure}[t]
\includegraphics[width=0.99\columnwidth]{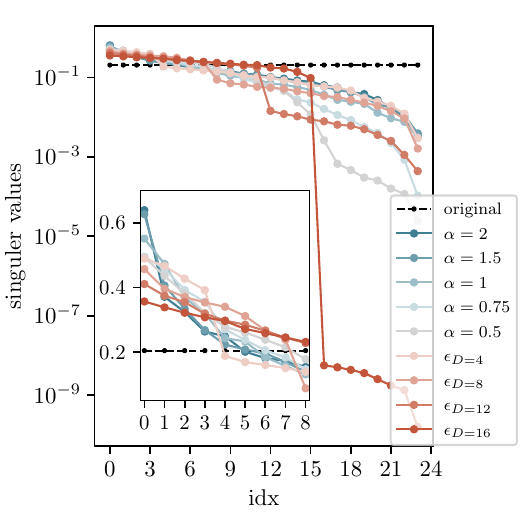}
\caption{\label{fig:disentangler} Singular values of $U\Theta$ after optimization with respect to $U$ using R\'enyi-$\alpha$ and $\epsilon_D$ as cost functions, where $\Theta$ is a random tensor.
The inset shows a zoom-in view.
}
\end{figure}

As an illustrative example, we consider the case when the tensor $\Theta_{(c,s_l),(s_r,b)}$ is randomly initialized and of size $(6,4,4,6)$.
The tensor $\Theta_{(c,s_l),(s_r,b)}$ has approximately constant singular values in the SVD carried out between the indices pair $(c,s_l),(s_r,b)$.
We benchmark on different losses with the optimal unitaries $U$ found by by minimizing R\'enyi-$\alpha$ entropy and the truncation error $\epsilon_{D}$.
The overall result is shown in Table~\ref{table: disentangler} and the singular values after the optimization is plotted in Fig.~\ref{fig:disentangler}.

We observe that all the truncation errors $\epsilon_{D}$ become smaller after the minimization for all different cost functions considered.
This suggest that utilizing the gauge degree of freedom and inserting the unitary is in general helpful regardless of cost function chosen.
Nevertheless, the resulting singular values depend on the cost function chosen and could have orders of magnitude difference in truncation error.

For choosing R\'enyi-$\alpha$ as cost function, we observe that the singular values modified varies ``smoothly".
For smaller $\alpha$ we see smaller singular values in the tail while for larger $\alpha$ we see the first few singular values have larger values.
This means minimizing R\'enyi-$\alpha$ entropy for a smaller $\alpha$ leads to better results for a larger truncated bond dimension $D$.
And minimizing a larger $\alpha$ leads to better results when the truncated bond dimension $D$ is smaller.
This is expected from the fact that for $\alpha\rightarrow\infty$, it corresponds to $\epsilon_{D=1}$.
There are crossings in between and the optimal $\alpha$ depends on the truncated bond dimension.
It is worth noting that, in all cases, minimizing the R\'enyi-$\alpha$ entropy leads to smallest R\'enyi-$\alpha$ entropy but could have larger truncation error comparing to the result of minimizing $\epsilon_D$.

In contrast to the ``smooth" change in singular values when minimizing the entanglement entropy, the singular values obtained from minimizing $\epsilon_D$ show sharp drop at the corresponding bond dimension $D$.
The optimization takes into account the information of the specific bond dimension $D$ and pushes down all singular values afterward to minimize the truncation error.
For a given bond dimension $D$, we see that minimizing $\epsilon_D$ always leads to the smallest truncation error $\epsilon_D$ even while it may have larger entanglement entropy.

The moral we learned from this illustrative example is that the conventional way of ``disentangling" a tensor $\Theta$ by minimizing entropy indeed brings down the truncation error, but (at least locally) in a sub-optimal way.
Disentangling modifies the overall spectrum but does not utilize the information about the anticipated truncated bond dimension $D$.
The direct minimization of truncation error utilizes such information to avoid the ambiguity in choosing $\alpha$ and thus could potentially lead to smaller truncation error.

In practice, the spectrum for a physical system would not be a constant but decay exponentially.
We observe similar behaviour and direct minimization of truncation error works slightly better than the disentangling approach (See Appendix~\ref{appendix:MM}).
However, the optimization problems are prone to get stuck at local minima when minimizing truncation error while in general minimizing entanglement entropy are more robust in general.
As a result, we report the results based on minimizing R\'enyi-$\frac{1}{2}$ entanglement entropy in this work.

To summarize, we have described two complementary approaches shifting the orthogonality hypersurface $\Psi^{[l]}=A^{[l]}\Psi$.
The variational MM treats the whole shifting as a variational problem and solves it with alternative least square update.
The sequential MM instead focuses on solving the optimal tripartite decompositions locally and build up the solution from these decompositions.
Diagrammatically, MM corresponds to going from Fig.~\ref{fig:MM_move}a to Fig.~\ref{fig:MM_move}b.
Going from Fig.~\ref{fig:MM_move}b to Fig.~\ref{fig:MM_move}c, we apply standard variational MPO-MPS compression.
Similar to the numerical algorithms in 1D, the move from Fig.~\ref{fig:MM_move}a to Fig.~\ref{fig:MM_move}c is the fundamental step.
The actual error for this complete move is the difference $\lVert \ket{\psi} - \ket{\psi'} \rVert^2$ for state before and after.
This is determined by the two columns before, $\Psi^{[l]}, B^{[l+1]}$, and the two columns after, $A^{[l]}, \Psi^{[l+1]}$, because the rest of the network can be contracted to identity.
In principle, we can again treat this as a variational problem and perform alternative least square update.
This step is more expensive and considered to be optional.
We include the complete sequential MM algorithm and describe other technical details in Appendix~\ref{appendix:MM}.

\begin{table*}[t]
  \centering
\begin{tabular}{ |p{2.cm}||p{1.5cm}|p{1.5cm}|p{1.5cm}|p{1.5cm}|p{1.5cm}|p{1.7cm}|p{1.7cm}|  }
 \hline
 \multicolumn{8}{|c|}{Comparison of disentangling approach and direct minimization of truncation error } \\
 \hline
\diagbox[linewidth=0.2pt, width={2.15cm}, height=0.6cm]{Minimize}{Result}
 & R\'enyi-2 & R\'enyi-1 & R\'enyi-$\frac{1}{2}$ & $\epsilon_{D=4}$ & $\epsilon_{D=8}$ & $\epsilon_{D=12}$ & $\epsilon_{D=16}$ \\
\hline
-- & 3.18& 3.18& 3.18& 0.833& 0.667& 0.5& 0.333\\
\hline
R\'enyi-2 & \textbf{1.56}& 2.09& 2.53& 0.28& 0.116& 0.0432& 0.0114\\
R\'enyi-1.5 & 1.58& 2.06& 2.5& 0.26& 0.1& 0.0364& 0.00945\\
R\'enyi-1 & 1.72& \textbf{2.04}& 2.41& 0.236& 0.0645& 0.0181& 0.00344\\
R\'enyi-0.75 & 1.86& 2.09& \textbf{2.35}& 0.266& 0.0608& 0.00727& 0.000507\\
R\'enyi-0.5 & 1.94& 2.16& 2.36& 0.32& 0.0727& 0.00656& 7.59e-06\\
\hline
$\epsilon_{D=4}$ & 1.8& 2.12& 2.5& \textbf{0.208}& 0.0964& 0.0373& 0.0108\\
$\epsilon_{D=8}$ & 2.02& 2.17& 2.43& 0.374& \textbf{0.0336}& 0.0133& 0.00407\\
$\epsilon_{D=12}$ & 2.26& 2.37& 2.47& 0.465& 0.165& \textbf{0.000714}& 0.000184\\
$\epsilon_{D=16}$ & 2.52& 2.62& 2.69& 0.561& 0.273& 0.087& \textbf{9.95e-17}\\
\hline
\end{tabular}
\caption{\label{table: disentangler} 
The resulting values for R\'enyi-$\alpha$ entropy and truncation error $\epsilon_D$ of $U\Theta$ after minimizing R\'enyi-$\alpha$ or $\epsilon_D$ as the cost function. The first row are values for the original tensor $\Theta$. Disentangling and direct minimizing truncation error both leads to smaller truncation error while direct minimizing truncation error gives smaller truncation error. The minimal value of each column is highlighted in boldface.
}
\end{table*}

\subsection{Discussion of isoTNSs}

To conclude the section of isoTNSs, we discuss and give some general remarks on the properties of isoTNSs and the related work.

Similar 2D isoTNS ans\"atze have also been proposed recently with different approaches for shifting the orthogonality hypersurface~\cite{haghshenas2019conversion,hyatt2019dmrg}. The isoTNSs can be generalized to higher dimensions or different lattice geometries, for example, the recent work on 3D cubic lattice~\cite{tepaske2020three}.

We note that it is possible for isoTNSs in 2D to have a different arrangement in the direction of isometries that does not look like Fig.~\ref{fig:isoTN2}d.
An example is when columns $A$, $B$ are not contracted to isometries pointing only toward the orthogonality center.
As also pointed out in~\cite{haghshenas2019conversion}, the identity maps formed by the left and the right boundary are sufficient but not necessary condition for an isoTNS to have an orthogonality center.
For our work, we choose the ``natural" arrangement in the sense that it is the direct generalization of MPSs by viewing each column of the TNSs as one site.
In~\cite{haghshenas2019conversion}, columns of TNSs are turned into layers of unitaries as in quantum circuits, which corresponds roughly to the pattern as in Fig.~\ref{fig:isoTN2}d. All the work so far~\cite{haghshenas2019conversion,zaletel2020isometric,hyatt2019dmrg,zhang2020stability} consider a similar arrangement of isometries for numerical convenience.

Aside from the previous attempts to generalize the canonical form of MPSs to PEPSs~\cite{perez2009canonical}, recently~\cite{evenbly2018gauge} proposes a different generalization of canonical form for cyclic tensor networks by gauge fixing.
The proposed weighted trace gauge condition (WTG)
requires the left and right boundary matrices
to be \emph{proportional} to identity $\mathbb{1}$ for all the virtual bonds.
\rewrite{}{We briefly review the definition of WTG in Appendix~{\ref{appendix: WTG}}.}
For acyclic networks, the WTG condition is equivalent to the canonical form condition.
However, for general cyclic networks, the WTG condition is a weaker condition than the isometric condition, since WTG condition is only a gauge fixing which does not change the overall states.
TNSs in canonical form defined with WTG is the same manifold of general TNSs.
In contrast, isoTNSs restricted the states and are sub-manifold of the general TNSs.
Notice that applying WTG to isoTNSs results in the $\Gamma-\Lambda$ form~\cite{vidal2003efficient,vidal2004efficient}.
The isometric condition is still satisfied by combining the bond tensors and the site tensors.

One distinguishing property of isoTNSs from TNSs is the absence of internal correlation. 
\rewrite{}{The direct renormalization and truncation for general cyclic TNSs are not optimal due to internal correlations.
One class of prominent examples are cyclic tensor networks with corner double line tensors\mbox{~\cite{gu2009tensor,evenbly2015tensor}}.
The cycle entropy $S_{\mathrm{cycle}}$ on a bond is a measure defined in~{\cite{evenbly2018gauge}} to quantify this physically redundant information contained in cyclic networks, i.e., internal correlations.
We give a brief review of the definition of the cycle entropy $S_{\mathrm{cycle}}$ in Appendix~{\ref{appendix: WTG}}.
When $S_{\mathrm{cycle}} = 0$, the bond does not carry physically redundant information.
However, TNSs usually have non-zero internal correlation, i.e., $S_{\mathrm{cycle}}\neq 0$.
One advantage of isoTNSs representation is that the cycle entropy $S_{\mathrm{cycle}}$ is always zero for all the bonds by construction.
This is a property following the definition of the cycle entropy $S_{\mathrm{cycle}}$.
While we discuss this in more detail in Appendix~{\ref{appendix: WTG}}, we provide an intuitive argument here.
IsoTNSs are states generated under sequential unitaries from the product states~\mbox{\cite{schon2005sequential,banuls2008sequentially}}.
The bonds in isoTNSs correspond to the actual physical degree of freedom, on which the unitaries act, and hence, $S_{\mathrm{cycle}} = 0$.
}

\section{\label{sec:Algorithm} Algorithm for Isometric TNS}

We describe two types of algorithms for isoTNSs, namely time evolution algorithm $(\text{TEBD}^2)$ and ground state search algorithm by variational minimization of energy $(\text{DMRG}^2)$.
We formulate both algorithms as the minimization problems of certain cost functions for local tensors.
In this fashion, both algorithms are iterative algorithms performing local updates over each tensor.
Here we consider the case where the tensor updated is always at the orthogonality center by shifting the orthogonality hypersurface.

\subsection{ $\mathrm{TEBD}^2$ algorithm}

The TEBD algorithm with MPSs is an algorithm utilizing local updates to perform time evolution.
The TEBD-like algorithms for time evolution consist of three parts:
(i) Suzuki-Trotter decomposition of the time evolution operator $\hat{U}(dt) = \prod_{i} e^{-idt H_i}$ of local Hamiltonian $H=\sum_i H_i$ into a set of two-site local operators
(ii) local updates by applying the time evolution operator following optimal bond truncation at the orthogonality center
(iii) shifting of orthogonality center.
Combining these with 1D isoTNSs, i.e. MPSs, the resulting TEBD-like algorithms are similar algorithms slightly varying in the implementation details~\cite{white2004real,daley2004time,vidal2003efficient,vidal2004efficient}.

\begin{figure}[t]
\includegraphics[width=1.\columnwidth]{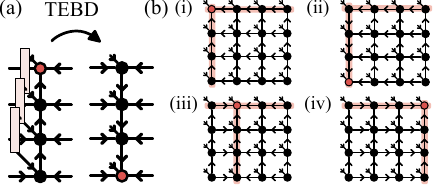}
\caption{\label{fig:algorithm-tebd}
The $\text{TEBD}^2$ algorithm:
(a) The 1D TEBD 
(b)(i) Beginning with isometries pointing toward the upper left, one TEBD step as in (a) on the column would bring the orthogonality center down as in (ii).
The MM as in Fig.~\ref{fig:MM_move} bring the arrow back up and shift the orthogonality hypersurface as in (iii).
Repeating this through the system, we arrive at the form as in (iv) in the end.
With a anti-clockwise $90^\circ $ rotation of (iv) we are back to (i) with column and row reverse. We then repeat the same steps and evolve the rows.
}
\end{figure}

The $\text{TEBD}^2$ algorithm is the 2D generalization of the TEBD algorithm with 2D isoTNSs.
The three building blocks work similarly in 2D.
For part (i), we consider the Suzuki-Trotter decomposition of the time evolution operator $ \hat{U}(dt) = \prod_{r,i} e^{-idt H^r_i} \prod_{c,j} e^{-idt H^c_j}$, where $H^r_i$ and $H^c_j$ are the terms of the local Hamiltonian acting on columns and rows.
For part (ii), we perform similar local updates at the orthogonality center.
Since optimal bond truncation is guaranteed by the isometric form, there is no difference to the 1D algorithm except the additional indices.
However, the computational cost can be drastically reduce from $\mathcal{O}(D^9)$ to $\mathcal{O}(D^5)$ by applying the two-site gate update on the reduced tensors~\cite{lubasch2014algorithms}.
Moreover, the reduced tensor update is optimal for isoTNSs, which is different to the general TNSs~\cite{PhysRevB.81.165104,lubasch2014algorithms}.
For part (iii), we utilize the SVD and MM to move around the orthogonality center and orthogonality hypersurface.

We sketch the outline of $\text{TEBD}^2$ algorithm here. 
\begin{enumerate}[label=(\roman*)]
    \item Start with an isoTNS with all isometries pointing toward the top left as in Fig.~\ref{fig:algorithm-tebd}b(i).
    
    \item Perform 1D TEBD with reduced tensor update on the column of the orthogonality center. After the sweep, the isometries all points down as in Fig.~\ref{fig:algorithm-tebd}b(ii). 
    
    \item Perform MM to bring the orthogonality center forward to the next column Fig.~\ref{fig:algorithm-tebd}b(iii), then repeat the 1D TEBD as in step (ii).
    Continue and repeat this steps over all columns.
    
    \item The isometries now point toward the top right as in Fig.~\ref{fig:algorithm-tebd}b(iv).
    The orientation of the isometries has effectively been rotated by $90^\circ$ counterclockwise from the starting point, Fig.~\ref{fig:algorithm-tebd}b(i). We may thus go back to step (i) by rotating the network by $90^\circ$, exchanging the role of rows and columns, and repeat.
\end{enumerate}

Each $\text{TEBD}^2$ step evolves the system $dt$ with two cycles of operation from step (i) to (iv).
This is because on the first round we finish the application of all the terms on columns $ \hat{U}^\text{col}(dt) = \prod_{c,j} e^{-idt H^c_j}$ to the state.
With the rotation at step (iv), we interchange the columns and rows.
On the second round, we thus evolve the ``rows" of the original lattice.
After two rounds, we arrive at a $180^\circ$ rotated lattice with the system evolved by $\hat{U}(dt)$.
By repeating this, one can perform real- or imaginary-time evolution with isoTNSs. 
We note that while each 1D TEBD step is individually a 1st-order Trotterization, after four-rounds we obtain a 2nd-order Trotterization within columns and rows, as the effective reversal by $180^\circ$ during rounds 3-4 cancels out errors via symmetrization.
By using half time steps in the first and last time steps of the column updates, we could make the overall algorithm 2nd-order~\cite{zaletel2020isometric}.
The method is termed $\mathrm{TEBD}^2$ since it is a nested loop of the 1D TEBD algorithm~\cite{zaletel2020isometric}. 

The $\mathrm{TEBD}^2$ algorithm differs from the time evolution algorithm of general 2D TNSs~\cite{murg2007variational,lubasch2014algorithms} in the step of tensor updates.
In general, application of the time evolution operator increases the bond dimensions of the TNSs.
For isoTNSs, we can simply truncate the bond dimension by a local SVD, which is the \emph{globally} optimal truncation because of the isometry conditions.
For generic TNSs, local truncation is not optimal because it does not take into account the information of the rest of the tensor network.
Instead, one has to solve the minimization problem approximating the time-evolved state $\ket{\psi(t+dt)}$,
\begin{equation}
    \argmin_{x}\  \Big \lVert \ket{\phi} - \hat{U}(dt) \ket{\psi(t)} \Big \rVert^2 .
\end{equation}
The $\ket{\psi(t)}$ denotes the original state at time $t$.
One updates a single local tensor $x$ in $\ket{\phi}$, which is a 2D TNS of same fixed bond dimensions as the original TNS~\footnote{In practice, one update the reduced tensor to lower the computational cost~\cite{lubasch2014algorithms}.}.
The optimal update is given by solving the systems of linear of equation with the norm matrix $N$ and vector $b$, similar to the problem described in variational MM.
For TNSs, the evaluation of the norm matrix $N$ and vector $b$ involves the approximate TNSs contraction for the environment, which gives rise to the difference in simple and full update scheme~\cite{lubasch2014algorithms}.
The isometric condition provides the optimal truncation and avoids the need of solving the systems of equation and environment approximation.
This advantage comes with the cost of the truncation error in MM.
Nevertheless, the computational complexity for the time evolution algorithm decreases from $D^{10}$ for general TNSs to $D^7$ for isoTNSs, where the $D^7$ complexity comes from the MM (See Appendix~\ref{appendix:MM}).

The 1D TEBD algorithm with MPSs has two sources of error: the Trotterization error $\epsilon_\text{Trotter}$ and truncation error $\epsilon_\text{trunc}$ due to the restricted bond dimension.
Suppose we want to evolve the system to time $T$ with a controllable targeted error of order $\epsilon_T$, we could keep the error $\epsilon_\text{Trotter}$ and $\epsilon_\text{trunc}$ both around the order $\sim \epsilon_T$.
The truncation error $\epsilon_\text{trunc}$ could be made smaller than $\epsilon_T$ by increasing the bond dimension.
The Trotterization error can be made as small as one wish by decreasing step size $\delta t$ or increasing the order $p$ of Trotterization~\cite{paeckel2019time}.
More precisely, the Trotterization error of each time step $\delta t$ is $\sim \delta t^{(p+1)}$ in $\mathcal{L}_2$ norm and the accumulated Trotterization error is estimated by the sum of error at each time step and is about $\sim T {\delta t}^{p}$~\footnote{
For an error per unit time $\epsilon = \epsilon_T / T$ to be satisfied, first order method needs $\frac{T}{\delta t} = \frac{T}{\epsilon} $ time steps, while second order method $\frac{T}{\delta t} = \frac{T}{\epsilon^{1/2}}$ and so and so on. The total evaluation for first order TEBD is then $\frac{T}{\epsilon} \times 2$, and for second order $\frac{T}{\epsilon^{1/2}} \times 5$. Higher order is computational preferred if one aims at higher accuracy $\epsilon \ll 1$.
}.

The $\text{TEBD}^2$ algorithm has one additional source of error: the error in the MM $\epsilon_\text{MM}$.
This additional error affects the optimal choice of step size and order of Trotterization, depending on the use case.
Crucially, unlike in 1D, where the truncation per step $\epsilon_{\textrm{trunc}} \to 0$ as $\delta t \to 0$ (because the state does not change), in 2D $\epsilon_{\textrm{MM}}$ remains finite even as $\delta t \to 0$. This is because the MM is generally not exact (though $\epsilon_{\textrm{MM}} \to 0$ as $D \to \infty$).
To assess the consequence for real-time evolution, suppose we are interested in evolving a system to a specific total time $T$, and assume the MM error $\epsilon_\text{MM}$ is fixed, there exists an optimal step size $\delta t$ and order $p$, which minimizes the sum of Trotterization error $\epsilon_\text{Trotter}$ and MM error $\epsilon_\text{MM}$, i.e. $\epsilon_\text{total} = (T{\delta t}^{p})^2  + \frac{T}{\delta t} \epsilon_\text{MM}$.
By iterating through $p=1,2,\ldots$, we can solve the minimization problem given the values of $T$ and $\epsilon_\text{MM}$ and obtain the error $\epsilon_\text{total}^*$ and step size $\delta t^*$ for the corresponding $p$. 
Comparing the total error obtained for different $p$, we can pick the optimal $p^*$ and the corresponding optimal step size $\delta t^*$.
When considering imaginary-time evolution, this error $\epsilon_\text{total}$ acts against the deceasing of energy $\sim \epsilon_\text{E}e^{-\Delta E_\text{gap} d\tau}$ per time step $d\tau$. The converging results thus have an error in energy $\epsilon_\text{E} \sim a{\delta \tau}^{2p} + b\frac{\epsilon_\text{MM}}{\delta \tau}$~\cite{zaletel2020isometric}.

\subsection{ $\mathrm{DMRG}^2$ algorithm}

DMRG is a variational energy minimization algorithm for the ground state with MPSs as the variational ansatz.
Extending the algorithm for TNSs is considered in~\cite{verstraete2004renormalization,pivzorn2010fermionic} with the main drawbacks of high complexity $\mathcal{O}(D^{12})$ and numerical instability.
Here, we first review the general approach of energy minimization with TNSs and then discuss the difference when isoTNSs were used.
\rewrite{}{There is recent proposal on fixing this issue~{\cite{lee2022there}}, which shows potential to be applied to large system size in general.}  
The energy minimization algorithm with 2D isoTNSs are dubbed as $\text{DMRG}^2$ as it resembles the 1D DMRG algorithm.

In general, the energy minimization problem is solved by an iterative local update on each tensor $x$ is as follows,
\begin{equation}
    x^\text{update} \leftarrow \argmin_{x} \frac{\braket{\psi|\hat{H}|\psi}}{\braket{\psi|\psi}} .
\end{equation}
By introducing the Lagrangian multiplier $\lambda_E$,
\begin{equation}
\partial_{x^*}  \braket{\psi|\hat{H}|\psi} - 
\partial_{x^*} \lambda_E \braket{\psi|\psi} = 0 ,
\end{equation}
the solution of the optimization problem on a single tensor $x$ is given by the generalized eigenvalue problem
\begin{equation}
    H_{\mathrm{eff}} x = \lambda_E N x,
\end{equation}
where $H_{\mathrm{eff}}$ is the contraction of energy expectation value $\braket{\psi|\hat{H}|\psi}$ with leaving $x$ and $x^*$ tensors out.
And $N$ is the norm matrix, as defined in Sec.~\ref{subsec: isoTNS_2}, is the contraction of the norm $\braket{\psi|\psi}$ leaving the $x$ and $x^*$ tensors out.

\begin{figure}[t]
\includegraphics[width=0.95\columnwidth]{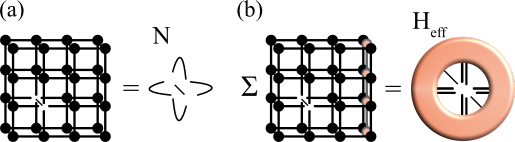}
\caption{\label{fig:algorithm-dmrg}
The generalization of $\text{DMRG}$ algorithm to 2D isoTNSs: (a) The norm matrix $N$ of isoTNSs is an identity operator $N=\mathbb{1}$ when the update site is the orthogonality center. (b) $H_\text{eff}$ is constructed by summing up the contraction of MPO representation of the Hamiltonian in the rows and columns.
}
\end{figure}

The crucial difference between considering TNSs and isoTNSs as variational ans\"atze is that the generalized eigenvalue problem reduces to standard eigenvalue problem with isoTNSs.
This is because the norm matrix of the orthogonality center is an identity operator $N=\mathbb{1}$ by the isometric condition. See Fig.~\ref{fig:algorithm-dmrg}.
This has the advantages of simplifying the computation and also stabilizing the algorithm, since the ill-conditioned generalized eigenvalue problem may return infinite or ill-disposed eigenvalues~\cite{pivzorn2010fermionic}.

In practice, our implementation expresses the Hamiltonian as a sum of 1D matrix-product operator (MPO) over the rows and columns. 
The expected energy is then a sum of the contraction over the isoTNS and MPOs.
Similarly, the $H_\text{eff}$ is a sum of the contraction over the isoTNS and MPOs while leaving the tensor $x$ and $x^*$ (See Fig.~\ref{fig:algorithm-dmrg}).
We contract each term approximately using the boundary MPSs approach~\cite{lubasch2014algorithms}.
\rewrite{}{The accuracy of the approximation is controlled by the bond dimension $D_\textrm{bMPS}$ of the boundary MPS and the overall cost of contracting the boundary MPS is $\mathcal{O}(D^6 D_\textrm{bMPS}^2 + D^4 D_\textrm{bMPS}^3)$.
In the following, we take $D_\textrm{bMPS} = 2 D^2$} and explicitly construct the matrix $H_{\mathrm{eff}}$ with the cost $\mathcal{O}(D^{12})$. 
Note that in principle only the matrix-vector multiplication $H_\text{eff}x$ operation is required for solving eigenvalue problem.
The complexity can be reduced to $\mathcal{O}(D^{10})$ if $H_{\mathrm{eff}}$ were not constructed explicitly.
It is possible that the approach above may be improved by the advanced optimal MPO compression scheme in~\cite{PhysRevB.102.035147}.

We sketch the outline of $\mathrm{DMRG}^{2}$ algorithm here. It is similar to the $\mathrm{TEBD}^2$ algorithm but replacing the local update with solving an eigenvalue problem.
\begin{enumerate}[label=(\roman*)]
    \item Start with an isoTNS with the orthogonality center at the left-most column.
    
    \item Perform the 1D DMRG over the column. That is we update each tensor $x$ in the column by solving the standard eigenvalue problem $ H_{\mathrm{eff}} x = \lambda x$ on the orthogonality center to obtain the lowest eigenvector $x^\text{update}$.
    We move the orthogonality center from site to site by SVD.
    
    \item Perform MM to bring the orthogonality center forward to the next column, then repeat the 1D DMRG as in step (ii).
    Continue and repeat this steps over all columns.
    
    \item At the end of the sweep, all tensors are updated. We perform a similar trick of rotation or a horizontal reflection to bring the isometries direction back to the starting arrangement as in step (i).
    
\end{enumerate}
The above steps give one $\mathrm{DMRG}^{2}$ sweep updating over all tensors.
The algorithm continues until the energy converges.

\rewrite{}{We would like to point out that the $\mathrm{DMRG}^{2}$ proposed here is not a standard variational algorithm, which optimize over the parameters of a single variational wavefunction. Instead, we optimize over a set of quantum states approximately connected by MM.
While we can variationally update the tensors in the column with orthogonality center to improve result, we always introduce an approximation (truncation) error when we move on with MM to optimize the tensors in the next column.
As a result, the variational energy does not monotonically decrease as in the standard DMRG algorithm, which is observed later in our numerical experiment and similarly in~Ref.~{\cite{hyatt2019dmrg}}.
The approximation of the ground state energy found by $\mathrm{DMRG}^{2}$ is thus, similar to $\mathrm{TEBD}^{2}$, bound by the MM error.
}

Another possible way to carry out $\mathrm{DMRG}^{2}$ with isoTNSs is to fix the isometric structure and not to perform MM when one sweeps through the lattice.
In that case, one has to compute the $N$ and solve for generalized eigenvalue problems.
One expects better condition numbers comparing to the case without gauge fixing.
With this approach one can study the representation power of the isoTNSs since there is no truncation involved.
Our observation is that the isometric condition itself without gauge fixing is still not stable.
Therefore, for the application in this paper, we consider the former approach instead.

\begin{figure}[h]
\includegraphics[width=1.\columnwidth]{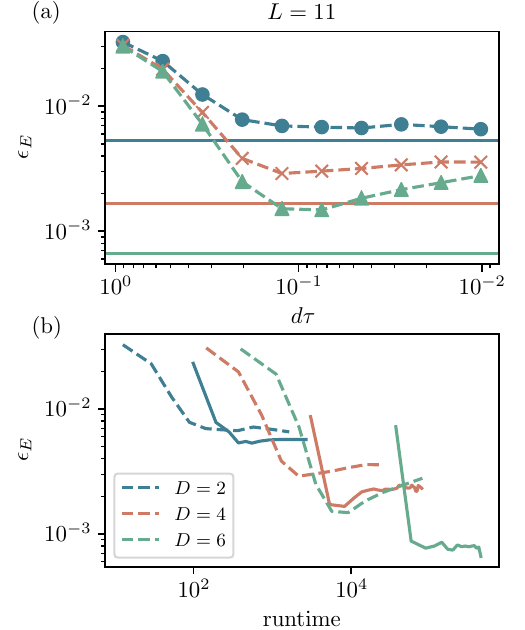}
\caption{\label{fig:algorithm-tebd-vs-dmrg}
The result from the $\text{TEBD}^2$ and $\text{DMRG}^2$ algorithms for the TFI model with $g=3.0$ on an $11\times 11$ square lattice.
(a) The relative energy error $\epsilon_E=(E-E_\text{exact})/\lvert E_\text{exact} \rvert$ is plotted as function of the Trotter step size $d\tau$.
We show data of different bond dimensions $D$ in different color.
The solid lines in the background represent error of $\text{DMRG}^2$ with the bond dimension $D$ of the same color.
(b) We plot the relative energy error $\epsilon_E$ against the runtime with $\text{TEBD}^2$ in dotted lines and $\text{DMRG}^2$ in solid lines.
The bond dimension on the orthogonality hypersurface are all $\eta=2D$ for $\text{TEBD}^2$ and $\eta=D$ for $\text{DMRG}^2$.
}
\end{figure}

As a demonstration of both algorithms discussed, we consider the transverse field Ising (TFI) model on the square lattice defined as
\begin{equation}\label{eq: TFI}
    H_{TFI} = -J\Big ( \sum_{\langle i,j \rangle} \hat{\sigma}^x_i \hat{\sigma}^x_j - g \sum_i \hat{\sigma}^z_i \Big ),
\end{equation}
where $\langle i,j \rangle$ denotes the nearest neighbors for site $i,j$.
We set $J=1$ as the unit.
In the thermodynamic limit, the TFI model exhibits a quantum phase transition from a symmetry-broken phase to a disordered phase at $g_c \approx 3.044$.
For benchmark, we consider a square lattice of size $L_x = L_y = 11$ with open boundary condition and $g=3.0$, close to the critical point.
We compare the ground state energy estimate obtained from the imaginary-time evolution using second-order $\text{TEBD}^2$ and $\text{DMRG}^2$ with the numerically exact results from 1D-DMRG simulation with bond dimension $D=1024$.
The result is plotted in Fig.~\ref{fig:algorithm-tebd-vs-dmrg}.
\rewrite{}{
The bond dimension of the overall isoTNS is denoted by $D$ and we allow the bond dimension in the orthogonality hypersurface to be $\eta$.
We consider the setup where $\eta=2D$ for $\text{TEBD}^2$ and $\eta=D$ for $\text{DMRG}^2$.}
\rewrite{The}{In Fig.~{\ref{fig:algorithm-tebd-vs-dmrg}}a, we show that for $\text{TEBD}^2$ the} energy estimates do not go down monotonically with the decrease of step $d\tau$ because of the MM error $\epsilon_\text{MM}$.
As described in previous section, the energy error could be fitted with $\Delta E_{\text{TEBD}^2} = a\epsilon_{MM}/d\tau + b d\tau^{2p}$ where $p$ is the order of Trotterization~\cite{zaletel2020isometric}.
The extrapolated optimal energy estimate is given by the minimum of the fit.
On the other hand, $\Delta E_{\text{DMRG}^2} \propto \epsilon_\text{MM}$.
We see in general $\text{DMRG}^2$ has smaller error estimates than the imaginary $\text{TEBD}^2$\rewrite{.}{ even when we use the same bond dimension $\eta=D$ over the orthogonality hypersurface.}
\rewrite{}{We plot the computational runtime in Fig.~{\ref{fig:algorithm-tebd-vs-dmrg}}b. Despite the difference in the scaling of computational complexity with respect to bond dimension, we observe that $\text{TEBD}^2$ and $\text{DMRG}^2$ reach similar accuracy at a given time for isoTNSs of different bond dimension. }

\begin{figure}[h]
\includegraphics[width=1.\columnwidth]{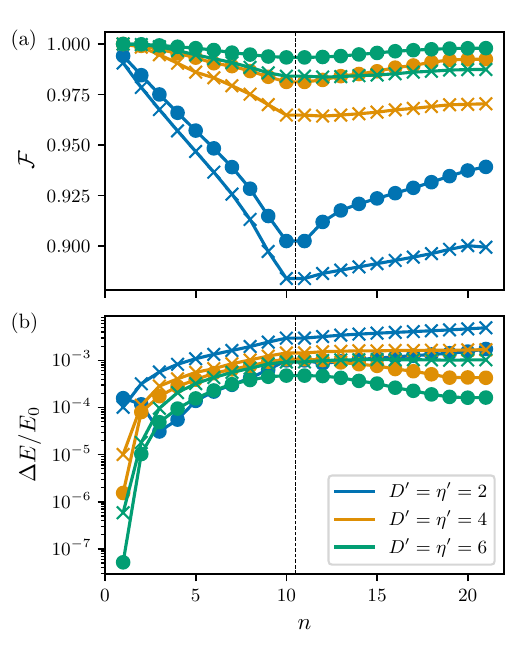}
\caption{\label{fig:test_mm}
Benchmark on repeating the MM over a $D=\eta=2$ isoTNS representing the ground state of the TFI model with $g=3.0$ on an $11\times 11$ square lattice.
(a) The fidelity $\mathcal{F}$ between the original state $\ket{\Psi_0}$ and the state $\ket{\Psi_n}$ after $n$ MM.
(b) The relative energy difference of the state $\ket{\Psi_n}$ comparing to the original state.
We perform in total $2L-1=21$ MM, sweeping from left to right and from right to left.
The MM are carried out in different bond dimensions $D', \eta'$ plotted in different color.
The solid circle and the cross mark the data obtained with variational MM and without variational MM.
The fidelity and the energy are measured by boundary MPS method with $D_\text{bMPS}=4\eta'^2$.
}
\end{figure}

\rewrite{}{
Let us finally comment on error made by repeated MMs. Note that we introduce an error $\epsilon_\text{MM}$ for each MM.
Consider a state $\ket{\Psi_0}$ that has the orthogonality center in the 0\textsuperscript{th}-column, by repeating the MM, we can move the orthogonality center to n\textsuperscript{th}-column--we denote the corresponding state as $\ket{\Psi_n}$.
The accumulated error of repeating the MM results in a deviation from the original state, which can be measured by the fidelity between the state $\ket{\Psi_n}$ and the original state $\ket{\Psi_0}$, i.e., $\mathcal{F} = \lvert \braket{\Psi_0 | \Psi_n}\rvert^2 \approx (1-\epsilon_\text{MM})^n$.
We take the $D=\eta=2$ isoTNS obtained from the DMRG\textsuperscript{2} for TFI model with $g=3.0$ on an $11 \times 11$ square lattice and perform repeatedly MM using bond dimension $D', \eta'$ sweeping from left to right and then from right to left. 
The result is shown in Fig.~{\ref{fig:test_mm}}.
We indeed observe a decrease of $\mathcal{F}$ with respect to $n$ in Fig.~{\ref{fig:test_mm}}a as we sweep from left to right.
However, we find that sweeping from the right end back to the original $0$\textsuperscript{th}-column does not cause further error in fidelity.
In fact, in some cases the fidelity $\mathcal{F}$ even increases.
We measure the corresponding energy for state $\ket{\Psi_n}$ and plot the error in Fig.~{\ref{fig:test_mm}}b.
The error density $\epsilon_\text{MM}$ introduces a bound on the accuracy that can reach by the algorithms.
We observe that continuing the left-right sweep of MM more than once lead to further degradation in fidelity.
But we also find, in some cases, an approximate fixed point where the fidelity almost converges with the continuing left-right sweep of MM. 
We show the data in Appendix~{\ref{appendix:data}}.
}

\section{\label{sec:Result} Spectral Functions}

As an application of the algorithms introduced above, we consider numerical evaluation of the dynamical spin structure factor (DSF), a.k.a. the spectral function.
The dynamical structure factor is defined as
\begin{equation}\label{DSF_def}
    S^{\alpha\alpha}({\bf k},\omega) =  \frac{1}{2\pi}  \sum_{\bf R} e^{-i{\bf k\cdot R}} \int_{0}^{\infty} 2\mathrm{Re} \Big [ e^{i\omega t} \mathcal{C}^{\alpha\alpha}({\bf R}, t) \Big ] dt,
\end{equation}
where the correlation functions $\mathcal{C}^{\alpha\alpha}({\bf R} ,t)=\braket{\hat{\sigma}^{\alpha\dagger}_{\bf R} (t) \hat{\sigma}^\alpha_{\bf 0}(0) }$ is evaluated with respect to the ground state $\ket{\psi_0}$.
It is of special importance since it gives us direct insight into the physical properties of the quasi-particles and the spectral properties of the Hamiltonian.
In addition, DSF can be measured by inelastic neutron scattering in experiment and can be computed using various methods  theoretically.

Here, we compute $S^{\alpha\alpha}$ numerically following the definition Eq.~\eqref{DSF_def}.
We first obtain the ground state $\ket{\psi_0}$ by $\mathrm{DMRG}^2$ with isoTNSs.
Then the locally perturbed state $\hat{\sigma}^\alpha \ket{\psi_0}$ is evolved using the $\mathrm{TEBD}^2$ algorithm.
Once we have the ground state $\ket{\psi_0}$ and the-time evolved state $e^{-i\hat{H}t}\hat{\sigma}^\alpha \ket{\psi_0}$, the time dependent real space correlation function $\braket{\hat{\sigma}^{\alpha\dagger}_{\bf R} (t) \hat{\sigma}^\alpha_{\bf 0}(0)}$ is obtained by the approximate contraction of TNSs.
For the data shown in next section, we apply linear prediction to double the time simulated from $T$ to $2T$.
In all cases, we multiply the data with the Gaussian ($\sigma_t\approx 0.44T$) which corresponds to a decay of factor $10$ at time $T$ and effectively smooth out and broaden the data in the frequency space.
Finally, the double Fourier transform of the correlation function gives us the spectral function.
We plot the spectral function in logarithmic scale with cutoff chosen to avoid showing the noise.

For the isoTNS with bond dimension chosen here, we observed MM error around $\epsilon_\text{MM} = \lVert \ket{\psi} - \ket{\psi'} \rVert^2 \sim 10^{-2}$ for sweeping the central column from the left to the right. The corresponding two error made for first-order and second-order $\text{TEBD}^2$ are similar.
For easier evaluation of time-dependent correlation function, we choose the first-order $\text{TEBD}^2$ method here~\footnote{With $\epsilon_\text{MM}\sim 0.01$ and $T\sim 1$, for first order method, we have optimal around $\delta t\sim 0.21$ and total error $\sim 0.09$. And for second order method, total error $\sim 0.05 $, $\delta t =0.4$.}.
We consider two different models to demonstrate the methods introduced could give qualitative insight into physical Hamiltonian.

\subsection{\label{subsec:TFI} Transverse field Ising model}

\begin{figure*}[t]
\includegraphics[width=0.99\textwidth]{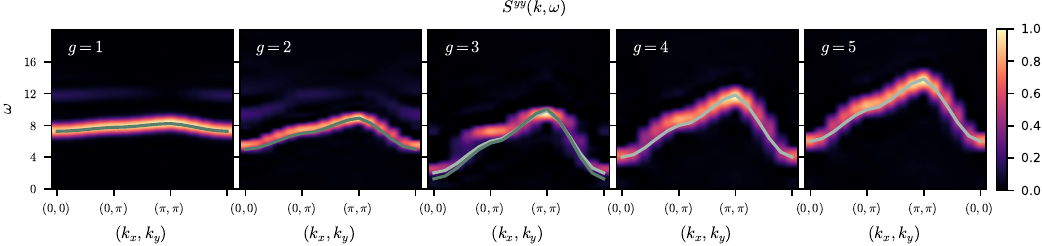}
\caption{\label{fig:TFI_DSF} The dynamical structure factor $S^{yy}({\bf k},\omega)$ for $g=1$ to $g=5$ in logarithmic color scale. The dark (light) green curve indicates the dispersion calculated perturbatively from the limit $g\ll 1$ ($g\gg1$). }
\end{figure*}

The TFI model, defined as in Eq.~\eqref{eq: TFI}, is a paradigmatic model for studying quantum many-body systems.
The ground state is ferromagnetically ordered for $g$ smaller than $g_c\approx 3.044$~\cite{rieger1999application,blote2002cluster}, and a disordered phase for $g>g_c$.
The excitation spectrum is known perturbatively in the large and small-$g$ limit by effective Hamiltonian method.
We compute the DSF $S^{{yy}}$ on a square lattice of size $L_x = L_y = 11$ for $g=1$ to $g=5$.
We plot the DSF result from the simulation and the perturbative calculation in Fig.~\ref{fig:TFI_DSF}.

In the limit $g\ll1$, the single-particle (magnon) excitation consists of a single spin flip costing energy $\sim 8J$.
To lowest non-vanishing order in $g$, we find a nearest-neighbor hopping model with energy, 
\begin{equation}
    \epsilon_{g\ll1} = 8 - \frac{g^2}{4}(1 + \cos(k_x) + \cos(k_y)) +\mathcal{O}(g^3) .
\end{equation}
The two-magnon excitations in the ferromagnetic phase form a bound state.
By simple counting, these bound states, which consist of two flipped spins on nearest neighbor sites, have energy $12J$ (lower than the two-particle continuum $\epsilon \sim 16J$).
We see for the $g=1$ and $g=2$ plots in Fig.~\ref{fig:TFI_DSF} the dispersion obtained by the simulation matches the result from perturbative calculation.
Moreover, we can see a slight signal of the bound states. However, throughout the full range of coupling, the two magnon continuum is not observed.

In the limit $g \gg 1$, by carrying out similar calculation, the energy of single-particle excitation is given as,
\begin{equation}
    \epsilon_{g\gg1} = g\Big[ 2- \frac{2}{g}(\cos(k_x) + \cos(k_y)) +\mathcal{O}(1/g^2) \Big ] .
\end{equation}
We again find that the DSF matches well with the perturbative calculation.

For $g=3$ near the critical point, we observe a small gap.
The gap size is slightly larger than the result in~\cite{vanderstraeten2019simulating} due to finite size effect.

\subsection{\label{subsec:Kitaev} Kitaev model on Honeycomb Lattice}

As a second example, we consider the Kitaev model on the honeycomb lattice~\cite{kitaev2006anyons}, consisting of three alternating spin couplings between bonds,
\begin{equation}
    H_\text{Kitaev} =  - J_x \sum_{\langle i j\rangle_x} \hat{\sigma}_i^x \hat{\sigma}_j^x - J_y \sum_{\langle i j\rangle_y} \hat{\sigma}_i^y \hat{\sigma}_j^y - J_z \sum_{\langle i j\rangle_z} \hat{\sigma}_i^z \hat{\sigma}_j^z
\end{equation}
The Kitaev model is an exactly solvable model describing two types of quantum spin liquids depending on the couplings.
The system is either a gapped $Z_2$ spin liquid with abelian excitations or a spin liquid with gapless Majorana and gapped flux excitations.
Here, we consider the isotropic coupling $J_x=J_y=J_z$, which belongs to the latter category.

The Kitaev model is of special importance for the reason that there are few examples of excitations of topological states that can be solved analytically.
Utilizing the Majorana fermions representation, one can obtain not only the ground state properties but also the properties for excitations.
The exact solutions for Kitaev model provided by \cite{knolle2014dynamics,knolle2015dynamics} for infinite system and \cite{zschocke2015physical} for finite system serve as a challenging benchmark for numerical simulation of the DSF for 2-dimensional systems.

\begin{figure*}[th!]
\includegraphics[width=1.\textwidth]{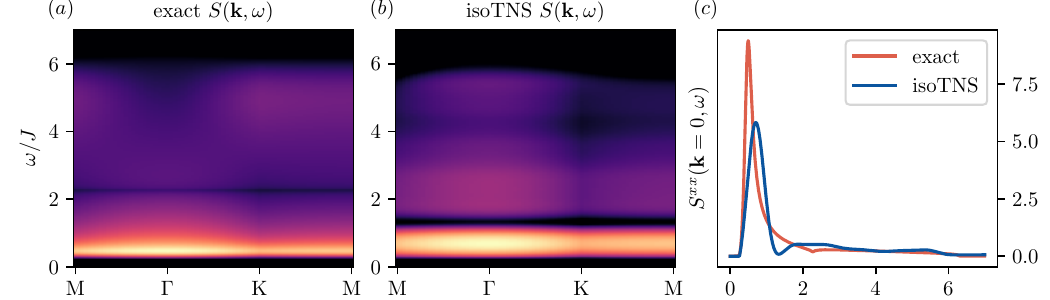}
\label{fig:DSF_kitaev}
\caption{\label{fig:Kitaev_DSF}  
The total dynamical structure factor $S({\bf k},\omega)=\sum_{\alpha} S^{\alpha\alpha}({\bf k},\omega)$ for Kitaev honeycomb model at the isotropic point in logarithmic color scale. (a) Exact result obtained for infinite system size following~\cite{knolle2014dynamics}. (b) Our numerical result obtained from simulation of system size $11 \times 11$. For better comparison, we perform smoothing by linear interpolation between discrete k-points. (c) The dynamical structure factor at $
\Gamma$ point, i.e. $S^{xx}_{{
\bf k}=0, \omega}$.
}
\end{figure*}

We obtain the DSF by a similar procedure with isoTNSs as before and plot the result in Fig.~\ref{fig:Kitaev_DSF}.
We compare the the data with the exact solution~\cite{knolle2014dynamics}.
The DSF at the isotropic gapless point is gapped due to the flux excitations and has a broad excitation continuum, as seen in Fig.~\ref{fig:Kitaev_DSF}a.
We see indeed in Fig.~\ref{fig:Kitaev_DSF}b, the simulation  reproduces the gapped excitation and broad dispersiveless signal due to fractionalization, similar to the analytic result.
In Fig.~\ref{fig:Kitaev_DSF}c, we examine the $S^{xx}(k=0, \omega)$ more closely, and confirm the excitation to be gapped.
While it is promising to see that we can qualitatively reproduce the result as from analytic solution, we would like to point out the result is still severely limited by the accuracy from both $\text{DMRG}^2$ and $\text{TEBD}^2$ method. See Appendix~\ref{appendix:data} for more details.

\section{\label{sec: Conclusion} Conclusion}

We introduced and discussed several properties of isoTNSs.
IsoTNSs are the natural generalization of MPSs in the isometric form to higher dimensions.
The isoTNSs in higher dimensions embed an effective 1D sub-region resembling the MPSs. 
This MPS-like sub-region is dubbed the orthogonality hypersurface and the known 1D algorithms can run efficiently within it.
We considered two different algorithms for shifting orthogonality hypersurface, which are the analogy to the orthogonal matrix decompositions in 1D.
We sketch the algorithms for time evolution and variational energy minimization with 2D isoTNSs.
And we demonstrate that one can efficiently simulate the real-time evolution for 2D systems and compute the dynamical structure factors of the TFI model on a square lattice and the Kitaev model on the honeycomb lattice.

The study of isoTNSs is related to quantum computation~\cite{slattery2021,PhysRevLett.128.010607}.
Essentially, isoTNSs are sequential and geometrically-local circuit ansatz.
IsoTNSs, in this perspective, are states that could be directly prepared on quantum computers.
The study of the properties, e.g.,variational power, of isoTNSs tells us the properties of the constant depth sequential quantum circuit~\cite{schon2005sequential,banuls2008sequentially} constructed by local gates of size growing logarithmically with the bond dimensions.

Algorithms for isoTNSs can be viewed as classical simulation algorithms for quantum circuits.
The insight from the study of isoTNSs could potentially leads to new quantum algorithms.
As we see, MMs are approximate algorithms for changing the isometric pattern in isoTNSs. In other words, they are approximate algorithms for re-ordering the quantum gates in the circuit.
They have potential applications in quantum state preparation and quantum circuit compilation. One example is that applying MM iteratively on MPSs yields 1D quantum circuits.
Additionally, it is shown that noisy quantum computers could be simulated efficiently classically by MPS~\cite{PhysRevX.10.041038}.
As a generalization of MPS, 2D isoTNSs could possibly yield a better classical simulation algorithm for 2D quantum circuits.

The accuracy achieved by isoTNSs algorithm could be improved by improving the algorithms for shifting orthogonality hypersurface.
A bottom-up approach would be to develop better tripartite decomposition to directly targeting the truncation error and overcoming the difficulties in the optimization.
Alternatively, one may consider the top-down approach which optimizes the variational Moses move using different gradient-based optimization.
This global update approach may result in a better minimum than the current local update approach solving the alternating least square problem.

One promising application of isoTNSs would be to combine it with Monte Carlo methods for studying ground states and time evolution.
IsoTNSs fit in Monte Carlo methods because isoTNSs allow ancestral sampling along the direction of causality, i.e. reverse direction of the arrows in the isometries.
This requires only single layer TNSs contraction which is shown can be contracted at a cheaper cost $\mathcal{O}(D^6)$~\cite{liu2017gradient,liu2019accurate}.
Samples from ancestral sampling are independent and do not have the problem with the auto-correlation time as in Markov Chain Monte Carlo (MCMC) sampling.
Therefore, it may be more efficient in terms of the number of Monte Carlo sweeps and the number of samples $N_{MC}$ comparing with the approach using general TNSs~\cite{liu2017gradient,liu2019accurate}.
In this approach, there is no truncation error since the orthogonality hypersurface is held fixed and it would serve as a good test for the variational power for isoTNSs.
Recently, the similar idea has been applied to general TNSs with the combination of importance sampling~\cite{vieijra2021direct}.

\begin{acknowledgments}
    \noindent We thank J. Knolle for providing data for Kitaev model.
    S. L. would like to thank Yantao Wu, Sajant Anand, Glen Evenbly, Laurens Vanderstraeten and Jheng-Wei Li for useful comments.
	\noindent This work was supported by the European Research Council (ERC) under the European Union's Horizon 2020 research and innovation program (grant agreement No. 771537). %
	F.P. acknowledges the support of the Deutsche Forschungsgemeinschaft (DFG, German Research Foundation) under Germany's Excellence Strategy EXC-2111-390814868. 
	S.L. and F.P. were supported by the DFG TRR80. The research is part of the Munich Quantum Valley, which is supported by the Bavarian state government with funds from the Hightech Agenda Bayern Plus.
\end{acknowledgments}

\appendix

\section{Optimization of isometries \label{appendix:opt_iso}}

Here, we provide an overview of the optimization problems with isometries.
In the most general form, the problem we are concerned with is the following.
Given $W \in C^{m\times n}, m\geq n$ and $f: W\rightarrow f(W)\in \mathbb{R}$, we want to find the optimal $W^\text{opt}$ leading to extreme of $f(W^\text{opt})$ under the isometry constraint $W^\dagger W = \mathbb{1}$.
These problems show up commonly in algorithms for isometric tensor networks and quantum circuits.

The simplest case for this type of problem is when $f$ is a linear function and with $W$ restricted to be real-valued, i.e. orthonormal matrix.
The problem is known as the orthogonal Procrustes problem and has close-formed solutions.
We review the solution and the proof of this type of problem in Appendix~\ref{appendix:polar}.
For general cases where $f$ is a non-linear function, one could consider to linearize the function and update $W$ in a similar fashion as in the linear case.
This is also known as the Evenbly-Vidal algorithm, which we review in Appendix~\ref{appendix:G-V}.
However, such an algorithm does not converge to the extrema in general.
As a result, we review the standard gradient descent methods over isometries~\cite{smith1994optimization} used in our previous work~\cite{zaletel2020isometric} and this work in Appendix~\ref{appendix:RGD}.

\subsection{Orthogonal Procrustes problem \label{appendix:polar}}

The orthogonal Procrustes problem~\cite{gower2004procrustes} is an optimization problem of finding the orthonormal matrix $W\in\mathbb{R}^{m\times m}$ which best transforms matrix $A\in \mathbb{R}^{l\times m}$ to matrix $B\in \mathbb{R}^{l\times m}$, that is
\begin{equation*}
    \argmin_W \lVert AW - B \rVert_F
\end{equation*}
Expanding out the expression, the problem is equivalent to
\begin{equation*}
    \argmax_W \mathrm{Tr}[WM]
\end{equation*}
where $M=B^\dagger A$ and $M\in \mathbb{R}^{m\times m}$.

More generally speaking, the optimization problem could be stated as finding the maxima of the function $f^\text{linear}: W\rightarrow f^\text{linear}(W)\in \mathbb{R}$, where $f^\text{linear}$ is a function linear in $W$.
Such a problem permits an exact solution. We first derive the maximum value of the function $f$, and show the solution which gives the maximum value.

We first find out the upper bound for the quantity $\mathrm{Tr}[WM]$.
Suppose the SVD of $M$ gives $M=USV^\dagger$,
\begin{align*}
    \mathrm{Tr}[WM] &= \mathrm{Tr}[WU\sqrt{S}\sqrt{S}V^\dagger] = \mathrm{Tr}[(\sqrt{S}U^\dagger W^\dagger)^\dagger (\sqrt{S}V^\dagger)] \\
    &=\braket{\sqrt{S}U^\dagger W^\dagger| \sqrt{S}V^\dagger}
\end{align*}
Since the matrix inner product induces the Frobenius norm. By the Cauchy-Schwarz inequality, we have
\begin{align*}
    \mathrm{Tr}[WM] &\leq \lVert\sqrt{S}U^\dagger W^\dagger\rVert_F  \lVert\sqrt{S}V^\dagger\rVert_F = \lVert\sqrt{S}\rVert_F  \lVert\sqrt{S}\rVert_F \\
    &= \mathrm{Tr}[S] .
\end{align*}
We use the invariance of the Frobenius norm under orthonormal transformation in the first equality.
The result suggest that the quantity $\mathrm{Tr}[WM]$ is upper bounded by $\mathrm{Tr}[S]$.
At the same time, we see that choosing orthonormal matrix $W=VU^\dagger$, we could have the maximum value $\mathrm{Tr}[S]$,
\begin{equation}
    \mathrm{Tr}[WM] = \mathrm{Tr}[VU^\dagger USV^\dagger] = \mathrm{Tr}[S]
\end{equation}
Therefore, the solution to the optimization problem is given by $W^\text{opt}=VU^\dagger$.

The generalized version of the problem consists of a matrix $M$ of dimension $(n, m)$, $m \geq n$, which could be complex-valued, $M\in \mathbb{C}^{n\times m}$.
Instead of optimizing over an orthonormal matrix, we are now looking for an isometry $W \in \mathbb{C}^{m\times n}$ which maximizes the absolute value $\left \lvert \mathrm{Tr}[WM] \right \rvert$.
Note that this is equivalent to maximize $\textrm{Re}\Big[ \mathrm{Tr}[WM] \Big]$ since one can always absorb the phase factor inside the isometry.
A similar derivation from the above holds.
The solution is then given as $\tilde{W}^\text{opt} = \tilde{V}{U}^\dagger$ from the reduced SVD $M = US\tilde{V}^\dagger$, where ${U}, S, \tilde{V}^\dagger$ are of dimension $(n, n)$, $(n, n)$, $(n, m)$. $\tilde{V}$ is now isometry. That is $\tilde{V}^\dagger \tilde{V} = I$, and $\tilde{V} \tilde{V}^\dagger = P_n$.

Note that with a matrix $M$ of dimension $(n, m)$, $m > n$, the solution is an isometry of dimension $(m,n)$.
That is there is a fixed direction for the isometry tensor.
In some cases in the isoTNSs algorithms, we require the isometry tensor to be in a different direction.
To satisfy the isometric condition needed in the algorithm, one must first truncate the surrounding tensors to having dimensions $n=m$ and then solve for the unitary matrix $W$.

\subsection{Evenbly-Vidal algorithm \label{appendix:G-V}}

In general, the optimization problems with isometries are nonlinear, such as the disentangling problem in the tripartite decomposition or finding the ground state with MERA.
The optimization problem then is to find the isometry matrix $W\in\mathbb{C}^{m\times n},\ m\geq n$, which minimizes the function $f: W\rightarrow f(W)\in \mathbb{R}$.
There is no exact solution in general.
It was proposed by Evenbly and Vidal~\cite{evenbly2009algorithms,evenbly2014algorithms} to linearize the function $f(W)$ and apply the exact solution from the previous section as an iterative update.

The idea of linearizing the function is to keep all the tensors fixed except the one being optimized. One could rewrite the function as,
\begin{equation}
    f(W)=\textrm{Tr}[W E_W] + \text{constant}
\end{equation}
where $E_W$ is the environment tensor of $W$ which in general may also depend on $W$.
Evenbly and Vidal proposed to update the isometry $W\leftarrow {W}' = {V}{U}^\dagger$ by treating $E_W$ as if it were independent of $W$ and $E_W = U\Sigma V^\dagger$.
The algorithm continues iteratively until convergence.

The algorithm has been generalized to cases where the environment tensor cannot be written easily as a tensor network~\cite{luchnikov2020riemannian}.
Instead of obtaining the environment tensor $E_W$ by tensor-network contraction, one can compute the derivative with respect to $W$ i.e., $\frac{\partial f}{\partial W}$.
The algorithm consists of iterative update of the isometry $W\leftarrow {W}' = {V}{U}^\dagger$  until convergence, where $\frac{\partial f}{\partial W}= U\Sigma V^\dagger$.

This algorithm could be viewed as a first-order optimization algorithm with the connection given in \cite{hauru2020riemannian,luchnikov2020riemannian}.
The algorithm converges to the optimal point only for restricted cases.
The algorithm converges for a negative (positive) definite quadratic form which includes examples such as the entanglement renormalization~\cite{luchnikov2020riemannian} and minimizing R\'enyi-$\alpha$ entropy with $\alpha=2$.
It is observed that it converges also for cases with $\alpha>1$.

\subsection{Gradient descent algorithm \label{appendix:RGD}}

Gradient descent algorithms are iterative optimization algorithms finding the local minimum given a differentiable function $f$.
Assuming a Euclidean geometry, a simple version of the gradient descent algorithm update the parameters $W$ with,
\begin{equation*}
    W \leftarrow W' = W - \gamma \times \frac{\partial f}{\partial W^*}
\end{equation*}
where $\gamma$ is known as the step size and is determined by the line-search procedure or other prescribed procedures.

To apply gradient descent algorithm to problems with isometry constraint, one can consider to modify the update with projection, i.e.
\begin{align*}
    U\Sigma V^\dagger &= W - \gamma \times \frac{\partial f}{\partial W^*}\\
    W &\leftarrow UV^\dagger . 
\end{align*}

A better way to adapt to the isometry constraint is to consider the Riemannian optimization approach~\cite{smith1994optimization} on Stiefel manifold with Euclidean metric~\cite{absil2009optimization}.
Such approach has recently been reintroduced for isometric tensor network and quantum circuits~\cite{hauru2020riemannian,luchnikov2020riemannian}.
The gradient is defined as the projection of partial derivative $\Gamma_W = \frac{\partial f}{\partial W^*}$ onto the tangent space $T_W$ and is given by
\begin{equation}
\nabla f = \Gamma_W - \frac{1}{2} W(W^\dagger \Gamma_W + \Gamma_W^\dagger W)
\end{equation}
Then the update is given by moving in the gradient direction along the geodesics with step size $\gamma$,
\begin{equation}
    W \leftarrow e^{-\gamma \nabla f} W .
\end{equation}
One may also consider generalized approaches not following the geodesics but retraction and update the isometry $W$ by The Cayley transform.
These approaches are equivalent up to second order~\cite{abrudan2008steepest}.

In addition, one can combine the Riemannian gradient descent with various different first-order gradient-based optimization method~\cite{becigneul2018riemannian,li2020efficient,luchnikov2020riemannian}.
In this paper and in our previous work on isoTNSs~\cite{zaletel2020isometric}, we have used the Riemannian non-linear conjugate gradient algorithm~\cite{polak1969note}.

\section{Moses move \label{appendix:MM}}

\subsection{Sequential Moses move algorithm \label{appendix:MM_algorthim}}

The sequential Moses move (MM) algorithm is shown in Alg.~\ref{alg_MM}.
We choose the convention that the algorithm takes central column $\Psi^{[l]}$ with isometries pointing downward as input and returns left-normalized column $A^{[l]}$ and central column $\Psi$ with isometries pointing upward.
The tripartite decomposition over $\Psi_{a,b,c}$ involves an optimization problem finding optimal unitary $U$, which is solved by methods described in Appendix~\ref{appendix:opt_iso}.

\begin{algorithm}[h]
\SetKwInOut{Input}{Input}
\SetKwInOut{Output}{Output}
\SetKwFor{For}{for (}{) $\lbrace$}{$\rbrace$}
\SetKwFor{TriSplit}{Tripartite Split (}{)$\lbrace$}{$\rbrace$}
\SetKwFor{TriSplit}{Tripartite Split (}{)$\lbrace$}{$\rbrace$}
\SetAlgoLined
\Input{ central column $\Psi^{[l]}$, cost function $\mathcal{L}$, bond dimension $D$, central bond dimension $\eta$}
\Output{ left-normalized column $A^{[l]}$ and central column $\Psi$ minimizing $\left \lVert \Psi^{[l]}- A^{[l]}\Psi \right \rVert^2$ with error $\epsilon$ }

$\text{idx}=0, \epsilon_{0} = \text{inf}$\;

\For{ $\textrm{idx} < N_\mathrm{row}-1$ }{
  $\text{idx} = \text{idx} + 1$ \;
  
  Group the indices of $\Psi^{[l]}[\text{idx}]$ to form $\Psi_{a,b,c}$\;
  
  \TriSplit{ $\Psi_{a,b,c}$ }{
    (i) SVD on $\Psi_{(a),(b,c)}$\\
    \quad $C_{a,b,c} \simeq A_{a,s} \Lambda_{s,s} V_{s,b,c} = A_{a,s} \Theta_{s,b,c}$\\
    
    (ii) Split the index $s$ to $s_l, s_r$, s.t.\\
    \quad $|s_l| < D$, $|s_r| < D$\\
    \quad $C_{a,b,c}\simeq A_{a,(s_l,s_r)} \Theta_{(s_l,s_r),b,c}$\\
    
    (iii) Find unitary $U$ from\\
    \quad $  \argmin_{U \in \mathrm{unitary}} \mathcal{L}( U\Theta)$ \\
    
    (iv) Insert the identity $I=U^\dagger U$\\
    \quad $A \leftarrow AU^\dagger$ and $\Theta \leftarrow U\Theta$\\
    
    (v) SVD on $\Theta_{(s_l,c), (s_r,b)}$, s.t.\\
    \quad $\Theta_{s_l,s_r,b,c} \simeq V_{s_l,c,t} \Lambda_{t,t} Q_{t, s_r,b} = \Psi_{s_l,c,t} Q_{t, s_r,b}$\\
    \quad $|t| < \eta$
    
    (vi) Collect the tensors and error\\
    \quad $\Psi_{a,b,c} \approx A_{a,s_l,s_r} \Psi_{s_l,c,t} Q_{t,s_r,b}$\\
    \quad $A^{[l]}[\text{idx}]\leftarrow A,\  \Psi[\text{idx}]\leftarrow \Psi$\\
    \quad $\epsilon_\text{split} = \epsilon_\text{SVD-1} + \epsilon_\text{SVD-2}$
  
  }
  Absorb $\Psi$ to $\Psi^{[l]}[\text{idx}+1]$ \;
  $\epsilon += \epsilon_\text{split}$
 }
 SVD on $\Psi^{[l]}[N] \simeq A^{[l]}[N] \Psi[N]$ \\
 $\epsilon += \epsilon_\text{SVD}$
\caption{The sequential Moses move algorithm}\label{alg_MM}
\end{algorithm}

\subsection{Comparison of Moses moves \label{appendix: comparison_MM}}
We consider the similar test on truncation error for a 2-column physical wavefunction as in~\cite{zaletel2020isometric} for Moses moves $\Psi^{[1,2]}\rightarrow A^{[1]}\Psi^{[2]}$ minimizing two different cost functions, R\'enyi-$\frac{1}{2}$ and $\epsilon_\eta$.
The wavefunction considered is the ground state of the transverse field Ising model on a two-columns ladder of size $2\times 20$ obtained from DMRG. The coupling strength is set to be different between the horizontal and vertical bonds, i.e. $H=\sum_i g\sigma^x_i - \sum_{\langle i,j\rangle_h} J_h \sigma^{z}_i \sigma^{z}_j- \sum_{\langle i,j\rangle_v} J_v \sigma^{z}_i \sigma^{z}_j$, where $g=2.5, J_h=0.5, J_v=1.5$.

\begin{figure}[t]
\includegraphics[width=0.99\columnwidth]{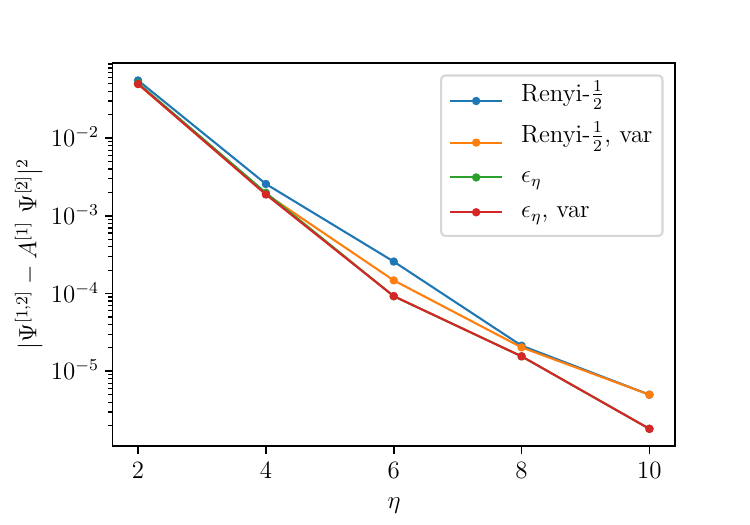}
\caption{\label{fig:dist_vs_trunc} 
Comparison of the error in the decomposition
$\left \lVert \Psi^{[1,2]} - A^{[1]}\Psi^{[2]}  \right \rVert^2 $
for the Moses Move with minimization based on R\'enyi-$\frac{1}{2}$ entropy and truncation error $\epsilon_\eta$. Furthermore, the variationally optimized solutions are also included. $\eta$ is the vertical bond dimension of $\Psi^{[2]}$. The test state $\Psi^{[1,2]}$ is the ground state of the TFI model on a two-columns ladder.
}
\end{figure}

The MM factorizes the 2-column wavefunction $\Psi^{[1,2]}$ into $A^{[1]} \Psi^{[2]}$, where $[1]$ denotes all physical indices for the first column and $[2]$ for the second column.
We consider the $A^{[1]}$ column to have fixed horizontal bond dimension $D_H$ and vertical bond dimension $D_V$ with $D_H= D_V = 2$, while for $\Psi^{[2]}$ column we have $D_H=2$ and $D_V=\eta$.
We compare the result of two different cost functions in the tripartite decomposition as described in Sec.\ref{subsec: MM} and measure the performance of MMs by $\left \lVert \Psi^{[1,2]} - A^{[1]}\Psi^{[2]}  \right \rVert^2 $ which depends on $\eta$ as shown in Fig.~\ref{fig:dist_vs_trunc}.
We see choosing the truncation error $\epsilon_{\eta}$ as a cost function gives slightly better results and the variational Moses move based on such initialization almost does not improve.

To understand the effect of cost functions in the tripartite decomposition, we take one tensor from the middle of the column during the MM and perform tripartite decomposition based on all different cost functions introduced in Sec.\ref{subsec: MM}.
We show the resulting truncation error and entanglement entropy in Table~\ref{table: dist_TFI}.
Similar to the observation in Sec.~\ref{subsec: MM}, disentangling and direct minimizing truncation error both leads to smaller truncation errors while direct minimizing truncation error gives a slightly better result.

\begin{table*}[t]
  \centering
\begin{tabular}{ |p{2.cm}||p{1.5cm}|p{1.5cm}|p{1.5cm}|p{1.5cm}|p{1.5cm}|p{1.7cm}|p{1.7cm}|  }
 \hline
 \multicolumn{8}{|c|}{Comparison of disentangling approach and direct minimization of truncation error } \\
 \hline
\diagbox[linewidth=0.2pt, width={2.15cm}, height=0.6cm]{Minimize}{Result}
 & R\'enyi-2 & R\'enyi-1 & R\'enyi-$\frac{1}{2}$ & $\epsilon_{\eta=2}$ & $\epsilon_{\eta=4}$ & $\epsilon_{\eta=6}$ & $\epsilon_{\eta=8}$ \\
\hline
-- & 0.386& 0.575& 0.914& 0.0198& 0.00263& 0.000178& 1.28e-05\\
\hline
R\'enyi-2 & \textbf{0.266} & 0.429& 0.749& 0.00751& 0.00103& 9.06e-05& 9.59e-06\\
R\'enyi-1.5 & \textbf{0.266} & 0.426& 0.74& 0.00673& 0.000973& 9.23e-05& 9.44e-06\\
R\'enyi-1 & 0.268& \textbf{0.425} & 0.727& 0.00558& 0.000943& 0.000105& 9.19e-06\\
R\'enyi-0.75 & 0.272& 0.426& 0.72& 0.00494& 0.000885& 0.000109& 9.22e-06\\
R\'enyi-0.5 & 0.278& 0.431& \textbf{0.718}& 0.00461& 0.000767& 7.49e-05& 9.88e-06\\
\hline
$\epsilon_{\eta=2}$ & 0.283& 0.435& 0.719& \textbf{0.00437}& 0.000623& 6.43e-05& 9.35e-06\\
$\epsilon_{\eta=4}$ & 0.289& 0.441& \textbf{0.717}& 0.00446& \textbf{0.000579}& 4.44e-05& 7.34e-06\\
$\epsilon_{\eta=6}$ & 0.296& 0.449& \textbf{0.718}& 0.00481& 0.000665& \textbf{1.99e-05}& 1.45e-06\\
$\epsilon_{\eta=8}$ & 0.298& 0.452& 0.725& 0.00511& 0.000842& 2.28e-05& \textbf{1.27e-06}\\
\hline
\end{tabular}
\caption{\label{table: dist_TFI} The result from a tensor taken in the middle of MM for the two-column wavefuncion described in Appendix~\ref{appendix: comparison_MM}. We compare the resulting values for R\'enyi-$\alpha$ entropy and truncation error $\epsilon_D$ of $U\Theta$ after minimizing R\'enyi-$\alpha$ or $\epsilon_D$ as the cost function.
Note that we choose a bond dimension $D_V=\eta$ for the column $\Psi^{[2]}$ so $\epsilon_D=\epsilon_\eta$.
The values in the first row are from the original tensor $\Theta$. Different to the test in Table~\ref{table: disentangler}, we see exponential decay in the truncation error $\epsilon_{\eta}$ for the original tensor in the first row. This indicates the singular values of the original tensor decay exponentially. We see utilizing the gauge degree of freedom to perform disentangling or direct minimizing truncation error could still lead to a substantial improvement in the truncation. The best values of each column are highlighted in boldface.}
\end{table*}

We would like to point out that the convergence results for MMs depend both on the cost function and the optimization procedure, because in general the optimization is not guaranteed to converge to the global minimum.
We consider two different types of the first-order Riemannian optimization methods: Riemannian Adam~\cite{li2020efficient} and Riemannian non-linear conjugate gradient method~\cite{polak1969note} with line search.
Setting R\'enyi-$\frac{1}{2}$ entanglement entropy as the cost function, the convergent result for the truncation error (thus the MM error) are similar with both optimization methods.
Setting the truncation error as the cost function, the Riemannian Adam gives slightly lower truncation error than that of having R\'enyi-$\frac{1}{2}$ entanglement entropy as cost function.
However, setting the truncation error as the cost function, the Riemannian conjugate gradient method with line search often gives worse result, i.e. higher truncation error, than the result of having R\'enyi-$\frac{1}{2}$ entanglement entropy as cost function.
Although we observe that the Riemannian Adam optimization with truncation error as cost function gives slightly better result, it also has slower convergence rate.
Moreover, it is sensitive to the step size and require problem-specific step size tuning.
As a result, we consider the Riemannian conjugate gradient methods with R\'enyi-$\frac{1}{2}$ entanglement entropy as cost function in this paper for efficiency reason.

\subsection{Implementation details \label{appendix:detail_MM} }

In practice, it is observed that increasing the bond dimensions on the orthogonality hypersurface could increase the representation power with less cost comparing to increase the bond dimensions uniformly.
As a result, we consider a maximal bond dimension $D$ throughout the tensor network, and a maximal bond dimension $\eta$ on the orthogonality hypersurface.
With this setup, Moses move would decompose the column $\Psi^{[l]}$ with bond dimension $\eta$ into two new columns $A^{[l]}$ and $\Psi$ with bond dimension $D$ and $\eta$ respectively.
See Fig.~\ref{fig:MM_move}.
The computation complexity of Moses move is $\mathcal{O}(\eta^3 D^4 + \eta^2 D^5)$ including the variational Moses move.

After the MM, one has to combine $\Psi$ column and $B^{[l+1]}$ column to form the new $\Psi^{[l+1]}$ column.
This step is similar to the standard MPO-MPS contraction. The direct contraction and truncation by randomized SVD have complexity $\mathcal{O}(\eta^2 D^5)$ and $\mathcal{O}(\eta^3 D^4)$ respectively.
Similar to MPSs compression, the one-sided truncation may lead to a sub-optimal result and one could consider it as initialization and further improve by variationally optimizing the truncated column with $\mathcal{O}(\eta^3 D^4)$.
Another possible way would be to consider combining the column variationally like variational MPO-MPS contraction, which gives the same structure in contraction as in variational MM but with now the single $\Psi^{[l+1]}$ column varying.
Thus, it would also have the same cost.
Note that direct contraction of $\Psi$ and $B^{[l]}$ following standard SVD truncation would however cost $\mathcal{O}(\eta^3 D^5)$, which should be avoided.
With the above counting, we show that shifting the columns from $\Psi^{[l]}B^{[l+1]}$ to $A^{[l]}\Psi^{[l+1]}$ has complexity $\mathcal{O}(\eta^3 D^4 + \eta^2 D^5)$ in general.

After MM and combining the columns, i.e. $\Psi^{[l]}B^{[l+1]}\rightarrow A^{[l]}\Psi^{[l+1]}$, an optional step to improve overlap can be considered by variationally maximizing $\braket{\Psi_\text{before}|\Psi_\text{after}} = \braket{\Psi^{[l]}B^{[l+1]} | A^{[l]}\Psi^{[l+1]} } $ again over the two new column $A^{[l]}$ and $\Psi^{[l+1]}$.
This variational optimization has contraction structure of four columns and the computation complexity $\mathcal{O}(\eta^3 D^4 + \eta^2 D^6)$.
In practice, we adapted all three variational procedures as in Fig.~\ref{fig:MM_move} when the numerical cost is acceptable.
Notice that even with the optional variational step, the overall computational complexity for the time evolution algorithm would still be cheaper than full update~\cite{lubasch2014algorithms} if the bond dimension in central column $\eta$ does not grow with $\mathcal{O}(D^2)$.

One can consider reducing the computational complexity further by decomposing the order-4 isometry into two trivalent tensors (omitting physical index).
Similar strategy is considered in the so-called triad network in the context of tensor network renormalization group~\cite{kadoh2019renormalization}.
The scaling could then be brought down to $\mathcal{O}(\eta^3 D^3)$ with the trade-off for less representation power. 
But it has the potential advantage of working with larger bond dimensions.

\section{Weighted trace gauge, internal correlation, and the corner double line tensors}
\label{appendix: WTG}

Here, we review the definition of weighted trace gauge (WTG) condition and cycle entropy $S_\text{cycle}$ quantifying the internal correlation introduced in~\cite{evenbly2018gauge}.
We consider a tensor network state $\ket{\psi}$, which in general includes bond matrices $\sigma$ on the virtual leg between two tensors.
The bond environment $\gamma^{ij}_{i'j'}$ is defined through the contraction of $\braket{\psi|\psi}$ leaving out the corresponding bond matrix $\sigma$ and its complex conjugation, where the indices $ij,i'j'$ are the corresponding bond indices.
The left and right boundary matrices are defined as $(\rho_L)^{i}_{i'} = \sum_{k,j,j'}\sigma_{kj} \sigma_{kj'}\gamma^{ij}_{i'j'}$, and $(\rho_R)^{j}_{j'} = \sum_{k,i,i'}\sigma_{ik} \sigma_{i'k}\gamma^{ij}_{i'j'}$, 
The WTG is the gauge choice over the bond such that the resulting left and right
boundary matrices $\rho_L$ and $\rho_R$ are proportional to the identity operator and the bond matrix is diagonal and positive and has elements in descending magnitude.
An algorithm to find the WTG is proposed in~\cite{evenbly2018gauge}.
For an acyclic tensor network, the WTG is equivalent to the standard canonical form.

For an acyclic tensor network, a bond is a ``bridge'' if by cutting the bond the tensor network becomes bipartite.
As a result, the bond environment factorizes, $\gamma^{ij}_{i'j'}=(\gamma_R)^{i}_{i'}(\gamma_L)^{j}_{j'}$, when the bond is a bridge.
To quantify the amount of internal correlation over a bond, the cycle entropy $S_\text{cycle}$ is defined as follows,
\begin{equation}
    S_\text{cycle} = -\sum_i \tilde{\lambda}_i \log \tilde{\lambda}_i
\end{equation}
where $\tilde{\lambda}_i = |\lambda_i| / (\sum_i |\lambda_i|)$ is the normalized eigenvalue of the transfer operator $(\sigma \otimes \sigma) \gamma$ formed by contracting the tensor product of the bond matrices to the bond environment.
The definition of cycle entropy is chosen such that it is gauge-invariant and is zero if the underlying bond is a bridge.
It is also invariant under the unitary transformation acting on the physical degree of freedom as this does not change the bond matrix $\sigma$ and the bond environment $\gamma$.
For an isoTNS, we see for any chosen bond, the corresponding bond environment always factorizes due to isometric conditions.
Therefore, isoTNSs have no internal correlation inside the tensor network.
The alternative way to see that isoTNSs have zero internal correlation is based on the property that the bond environment is invariant under unitary transformation acting on the physical degree of freedom.
The isoTNSs have zero cycle entropy because the product states have zero cycle entropy and isoTNSs are unitary transformation from product states.

One example of tensor network having internal correlation is TNS consisting of corner double line (CDL) tensors, which has the form
\begin{equation*}
\begin{diagram}
\draw (0,0) node (X) {};  
\draw[rounded corners] (1,2.5) rectangle (3,0.5);
\draw (1,1.75) -- (3.5,1.75);
\draw (1.75,0.5) -- (1.75,-0.5);
\draw (2.25,0.5) -- (2.25,-0.5);
\draw (1,1.25) edge[out=0,in=90] (1.75,0.5);
\draw (3,1.25) edge[out =180, in=90] (2.25, 0.5);
\draw (3., 1.25) -- (3.5, 1.25);
\draw (1., 1.25) -- (0.5, 1.25);
\draw (1., 1.75) -- (0.5, 1.75);
\end{diagram}\;      .
\end{equation*}
Each line in the tensor is a Kronecker-delta $\delta_{ij}$ of dimension $d$.
Taking this CDL tensor as an example, we will gain intuition on why isoTNS has no internal correlation.

The CDL tensor can be viewed as the left or right isometric form up to a normalization factor as shown below.
\begin{align*}
\applyTransferLeft{\bar{A}}{l}{A} &= \ \ d\ \times\ \drawMatrixLeft{l},\; \\ \nonumber \\
\applyTransferRight{A}{r}{\bar{A}} &= d \times \drawMatrixRight{r}\; .
\end{align*}

Consider a state defined by the CDL tensors in a loop over four sites as in~\cite{evenbly2018gauge}.
\begin{align}
&\ \ 
\begin{diagram}
\draw (0,0) node (X) {};  %
    \foreach \i in {0,...,3}
{
    \pgfmathsetmacro{\x}{(3*\i - 1) };
    \pgfmathsetmacro{\y}{(3*\i - 1) };
\draw[rounded corners] (\x+1,2.5) rectangle (\x+3,0.5);
\draw (\x+1,1.75) -- (\x+3.,1.75);
\draw (\x+3,1.75) -- (\x+3.5,1.75);
\draw (\x+3,1.25) -- (\x+3.5,1.25);
\draw (\x+0.5,1.75) -- (\x+1,1.75);
\draw (\x+0.5,1.25) -- (\x+1,1.25);
\draw (\x+1.75,0.5) -- (\x+1.75,-0.5);
\draw (\x+2.25,0.5) -- (\x+2.25,-0.5);
\draw (\x+1,1.25) edge[out=0,in=90] (\x+1.75,0.5);
\draw (\x+3,1.25) edge[out =180, in=90] (\x+2.25, 0.5);
}
\draw (-0.5, 1.75) edge[out=-180,in=-90] (-1.25, 2.5);
\draw (-1.25, 2.5) edge[out=90,in=-180] (-0.5, 3.25);
\draw (-0.5, 1.25) edge[out=-180,in=-90] (-1.75, 2.5);
\draw (-1.75, 2.5) edge[out=90,in=-180] (-0.5, 3.75);
\draw (12-0.5, 1.75) edge[out=0,in=-90] (12+.25, 2.5);
\draw (12+.25, 2.5) edge[out=90,in=0] (12-0.5, 3.25);
\draw (12-0.5, 1.25) edge[out=0,in=-90] (12+.75, 2.5);
\draw (12+.75, 2.5) edge[out=90,in=0] (12-0.5, 3.75);
\draw (-0.5, 3.25) -- (12-0.5, 3.25);
\draw (-0.5, 3.75) -- (12-0.5, 3.75);
\end{diagram}\; \label{eq: CDL_TN} \\ \nonumber \\
& \propto 
\begin{diagram}
\draw[double][->-=0.75] (0.5,1.5) -- (1,1.5); 
\draw[rounded corners] (1,2) rectangle (2,1);
\draw (1.5,1.5) node (X) {};
\draw[double][->-=0.75] (2,1.5) -- (3,1.5); 
\draw[rounded corners] (3,2) rectangle (4,1);
\draw (3.5,1.5) node {};
\draw[double][->-=0.75] (4,1.5) -- (5,1.5);
\draw[rounded corners] (5,2) rectangle (6,1);
\draw (5.5,1.5) node {};
\draw[double][->-=0.75] (6,1.5) -- (7,1.5); 
\draw[rounded corners] (7,2) rectangle (8,1);
\draw (7.5,1.5) node {};
\draw[double][->-=1.]  (8,1.5) -- (8.5,1.5); 
\draw[double][-<-=0.75]  (1.5,1) -- (1.5,.5);
\draw[double][-<-=0.75]  (3.5,1) -- (3.5,.5);
\draw[double][-<-=0.75]  (5.5,1) -- (5.5,.5);
\draw[double][-<-=0.75]  (7.5,1) -- (7.5,.5);
\draw (0.5, 1.5-0.03) edge[out=-180,in=-90] (0-0.03, 2.);
\draw (0-0.03, 2) edge[out=90,in=-180] (0.5, 2.5+0.03);
\draw (0.5, 1.5+0.03) edge[out=-180,in=-90] (0+0.03, 2.);
\draw (0+0.03, 2.) edge[out=90,in=-180] (0.5, 2.5-0.03);
\draw (8+0.5, 1.5+0.03) edge[out=0,in=-90] (9-0.03, 2.);
\draw (9-0.03, 2) edge[out=90,in=0] (8.5, 2.5-0.03);
\draw (8+0.5, 1.5-0.03) edge[out=0,in=-90] (9+0.03, 2.);
\draw (9+0.03, 2) edge[out=90,in=0] (8.5, 2.5+0.03);
\draw[double][-<-=0.5] (0.5, 2.5) -- (8.5, 2.5); .
\end{diagram}
\end{align}
The representation can be viewed as a periodic MPS or an generic 2D TNS. We can also view it as an isoTNS that has isometric direction forming loop and has no orthogonality center.
This is in fact an example of invalid isoTNS representation, which violates our assumption that the isometric direction of isoTNSs does not form loop and must have exactly one orthogonality center.
We will show in the following valid isoTNS representations of the same state, satisfying our assumption.

Firstly, we notice that the state defined in Eq.~\eqref{eq: CDL_TN} has redundancy in the tensor network representation and can be rewritten as
\begin{equation}
\begin{diagram}
\draw (0, 2.5) node (X) {};  %
    \foreach \i in {0,...,3}
{
    \pgfmathsetmacro{\x}{(3*\i - 1) };
    \pgfmathsetmacro{\y}{(3*\i - 1) };
\draw[rounded corners] (\x+1,2.5) rectangle (\x+3,0.5);
\draw (\x+3,1.25) -- (\x+3.5,1.25);
\draw (\x+0.5,1.25) -- (\x+1,1.25);
\draw (\x+1.75,0.5) -- (\x+1.75,-0.5);
\draw (\x+2.25,0.5) -- (\x+2.25,-0.5);
\draw (\x+1,1.25) edge[out=0,in=90] (\x+1.75,0.5);
\draw (\x+3,1.25) edge[out =180, in=90] (\x+2.25, 0.5);
}
\draw (-0.5, 1.25) edge[out=-180,in=-90] (-1.75, 2.5);
\draw (-1.75, 2.5) edge[out=90,in=-180] (-0.5, 3.75);
\draw (12-0.5, 1.25) edge[out=0,in=-90] (12+.75, 2.5);
\draw (12+.75, 2.5) edge[out=90,in=0] (12-0.5, 3.75);
\draw (-0.5, 3.75) -- (12-0.5, 3.75);
\end{diagram}\; \label{eq: CDL_TN_2} .
\end{equation}
The tensor network states in Eq.~\eqref{eq: CDL_TN} and in Eq.~\eqref{eq: CDL_TN_2} are equivalent but the tensor network in Eq.~\eqref{eq: CDL_TN} has bond dimension $d^2$ instead of $d$ due to the internal correlation.
When using generic TNS describing the state, this is often an issue and is hard to diagnose in tensor renormalization group~\cite{levin2007tensor} algorithms for generic TNS.
Recently, there are proposals in removing this redundancy locally~\cite{hauru2018renormalization,evenbly2018gauge}.

The tensor in Eq.~\eqref{eq: CDL_TN_2} can be normalized and identified as tensor in four different isometric forms:
\begin{align}
    \applyTransferLeftIso{\bar{A}}{l}{A} &= \ \ d\ \times\ \drawMatrixLeftIso{l}\; , \\ \nonumber \\
    \applyTransferRightIso{A}{r}{\bar{A}} &= d \times \drawMatrixRightIso{r}\; , \\ \nonumber \\
    \applyTransferMixedIso{A}{r}{\bar{A}} &= \drawMatrixMixedIso{r}\; , \label{eq: CDL_TN_2_boundary}  \\ \nonumber \\ 
    \applyTransferCenterIso{A}{r}{\bar{A}} \; &= d^2 \label{eq: CDL_TN_2_center}\;. 
\end{align}

Using the identities shown above, we can rewrite the state in Eq.~\eqref{eq: CDL_TN_2} as a valid isoTNS,
\begin{align}
& \ \
\begin{diagram}
\draw (0, 2.5) node (X) {};  %
    \foreach \i in {0,...,3}
{
    \pgfmathsetmacro{\x}{(3*\i - 1) };
    \pgfmathsetmacro{\y}{(3*\i - 1) };
\draw[rounded corners] (\x+1,2.5) rectangle (\x+3,0.5);
\draw (\x+3,1.25) -- (\x+3.5,1.25);
\draw (\x+0.5,1.25) -- (\x+1,1.25);
\draw (\x+1.75,0.5) -- (\x+1.75,-0.5);
\draw (\x+2.25,0.5) -- (\x+2.25,-0.5);
\draw (\x+1,1.25) edge[out=0,in=90] (\x+1.75,0.5);
\draw (\x+3,1.25) edge[out =180, in=90] (\x+2.25, 0.5);
}
\draw (-0.5, 1.25) edge[out=-180,in=-90] (-1.75, 2.5);
\draw (-1.75, 2.5) edge[out=90,in=-180] (-0.5, 3.75);
\draw (12-0.5, 1.25) edge[out=0,in=-90] (12+.75, 2.5);
\draw (12+.75, 2.5) edge[out=90,in=0] (12-0.5, 3.75);
\draw (-0.5, 3.75) -- (12-0.5, 3.75);
\end{diagram}\; \nonumber \\ \nonumber \\
&\propto
\begin{diagram}
\draw[->-=0.75] (0.5,1.5) -- (1,1.5); 
\draw[rounded corners] (1,2) rectangle (2,1);
\draw (1.5,1.5) node (X) {};
\draw[-<-=0.5] (2,1.5) -- (3,1.5); 
\draw[rounded corners] (3,2) rectangle (4,1);
\draw (3.5,1.5) node {};
\draw[-<-=0.5] (4,1.5) -- (5,1.5);
\draw[rounded corners] (5,2) rectangle (6,1);
\draw (5.5,1.5) node {};
\draw[-<-=0.5] (6,1.5) -- (7,1.5); 
\draw[rounded corners] (7,2) rectangle (8,1);
\draw (7.5,1.5) node {};
\draw[->-=0.75]  (8,1.5) -- (8.5,1.5); 
\draw[double][-<-=0.75]  (1.5,1) -- (1.5,.5);
\draw[double][-<-=0.75]  (3.5,1) -- (3.5,.5);
\draw[double][-<-=0.75]  (5.5,1) -- (5.5,.5);
\draw[double][-<-=0.75]  (7.5,1) -- (7.5,.5);
\draw (0.5, 1.5) edge[out=-180,in=-90] (0, 2.);
\draw (0, 2.) edge[out=90,in=-180] (0.5, 2.5);
\draw (8+0.5, 1.5) edge[out=0,in=-90] (9, 2.);
\draw (9, 2) edge[out=90,in=0] (8.5, 2.5);
\draw[-<-=0.5] (0.5, 2.5) -- (8.5, 2.5);
\end{diagram}\; \label{eq: CDL_iso1},
\end{align}
where the orthogonality center is on the first site as in Eq.~\eqref{eq: CDL_TN_2_center} and the tensor on the last site is as in Eq.~\eqref{eq: CDL_TN_2_boundary}.
Note that this is a valid isoTNS representation, i.e., the isoTNS has no loop in the isometric direction and has exactly one orthogonality center.
In the valid isoTNS representation of the state, there is no internal correlation.

The valid isoTNS representation of the state, however, is not unique.
In fact, we can also rewrite the tensor network in Eq.~\eqref{eq: CDL_TN_2} in the following form,
\begin{align}
& \ \
\begin{diagram}
\draw (0, 2.5) node (X) {};  %
    \foreach \i in {0,...,3}
{
    \pgfmathsetmacro{\x}{(3*\i - 1) };
    \pgfmathsetmacro{\y}{(3*\i - 1) };
\draw[rounded corners] (\x+1,2.5) rectangle (\x+3,0.5);
\ifthenelse{\i=3}{\draw (\x+1.75,1.75) edge[out=0,in=90] (\x+2.25, 1.25);}{\draw (\x+3,1.25) -- (\x+3.5,1.25);}
\ifthenelse{\i=0}{\draw (\x+2.25,1.75) edge[out =180, in=90] (\x+1.75, 1.25);}{\draw (\x+0.5,1.25) -- (\x+1,1.25);}
\draw (\x+1.75,0.5) -- (\x+1.75,-0.5);
\draw (\x+2.25,0.5) -- (\x+2.25,-0.5);
\ifthenelse{\i=0}{\draw (\x+1.75,0.5) -- (\x+1.75, 1.25);}{\draw (\x+1,1.25) edge[out=0,in=90] (\x+1.75,0.5);}
\ifthenelse{\i=3}{\draw (\x+2.25,0.5) -- (\x+2.25, 1.25);}{\draw (\x+3,1.25) edge[out =180, in=90] (\x+2.25, 0.5);} 
}
\draw (1.25, 1.75) -- (12-2.25, 1.75);
\end{diagram}\; \\ \nonumber \\
&\propto
\begin{diagram}
\draw[rounded corners] (1,2) rectangle (2,1);
\draw (1.5,1.5) node (X) {};
\draw[double][->-=0.75] (2,1.5) -- (3,1.5); 
\draw[rounded corners] (3,2) rectangle (4,1);
\draw (3.5,1.5) node {};
\draw[double][->-=0.75] (4,1.5) -- (5,1.5);
\draw[rounded corners] (5,2) rectangle (6,1);
\draw (5.5,1.5) node {};
\draw[double][->-=0.75] (6,1.5) -- (7,1.5); 
\draw[rounded corners] (7,2) rectangle (8,1);
\draw (7.5,1.5) node {};
\draw[double][-<-=0.75]  (1.5,1) -- (1.5,.5);
\draw[double][-<-=0.75]  (3.5,1) -- (3.5,.5);
\draw[double][-<-=0.75]  (5.5,1) -- (5.5,.5);
\draw[double][-<-=0.75]  (7.5,1) -- (7.5,.5);
\end{diagram}\; \label{eq: CDL_iso2},
\end{align}
which permit another valid isoTNS representation.
We notice that this isoTNS representation has bond dimension $d^2$, which comes from the price of encoding the correlation from the first site to the last site.

Although the isoTNS representation of the state is not unique and might subject to growth of bond dimension due to encoding long-range correlation, under this construction, it is not possible to add additional redundancy, i.e., internal correlation, to the valid isoTNS representation without violating the assumption or the isometric condition.
This is different from the generic tensor network as in Eq.~\eqref{eq: CDL_TN}, which in principle can have arbitrary bond dimension growth due to the internal correlation.

The state consisting of CDL tensors shows as an illustrative example as why isoTNS representation has zero cycle entropy.
It might be important to develop algorithm to understand and distinguish between the representation in Eq.~\eqref{eq: CDL_iso1} and that in Eq.~\eqref{eq: CDL_iso2}.

\section{Extra Data \label{appendix:data}}

\begin{figure}[h]
\includegraphics[width=1.\columnwidth]{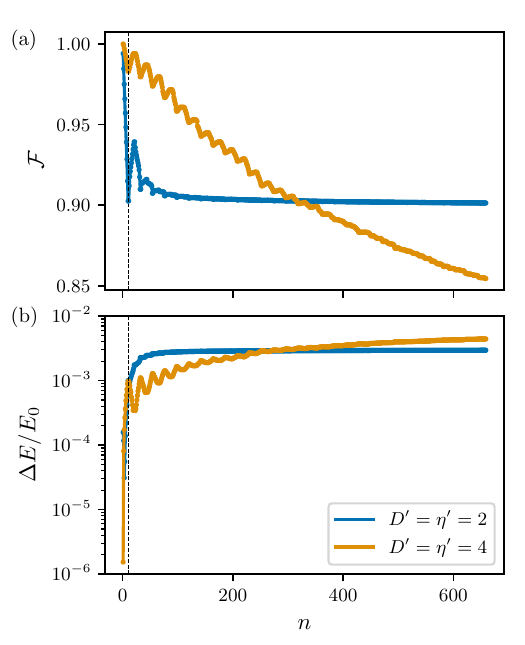}
\caption{\label{fig:test_mm_fp}
Benchmark on repeating the MM over a $D=\eta=2$ isoTNS representing the ground state of the TFI model with $g=3.0$ on an $11\times 11$ square lattice.
(a) The fidelity $\mathcal{F}$ between the original state $\ket{\Psi_0}$ and the state $\ket{\Psi_n}$ after $n$ MM.
(b) The relative energy difference of the state $\ket{\Psi_n}$ comparing to the original state.
We perform 30 left-right sweep, which is in total $30\times(2L-1)=630$ MM.
The MM are carried out in different bond dimensions $D', \eta'$ plotted in different color.
The solid circle and the cross mark the data obtained with variational MM and without variational MM.
The fidelity and the energy are measured by boundary MPS method with $D_\text{bMPS}=4\eta'^2$.
}
\end{figure}

In Fig.~\ref{fig:test_mm_fp}, we show the benchmark result on repeating the MM over a $D=\eta=2$ isoTNS representing the ground state of the TFI model with $g=3.0$ on an $11\times 11$ square lattice.
We observe that for MM using $D'=\eta'=2$, we can reach an approximate fixed point, where the fidelity almost remains the same.
Interestingly, for MM with $D'=\eta'=4$, we find better result for the initial sweeps but do not find an approximate fixed point.

\begin{figure}[h]
\includegraphics[width=1.\columnwidth]{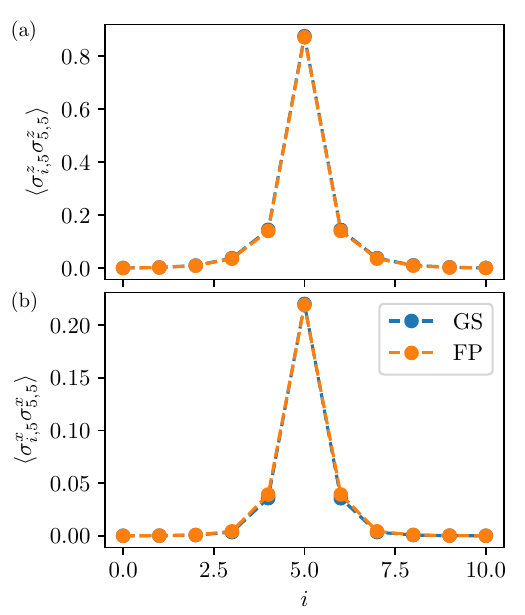}
\caption{\label{fig:fp_gs_corr}
The comparison of the connected correlation functions (a)  $\braket{\sigma^z_{i,5}\sigma^z_{5,5}}_{\textrm{c}}$ and (b) $\braket{\sigma^x_{i,5}\sigma^x_{5,5}}_{\textrm{c}}$.
The approximate fixed point wavefunction (FP) of $D'=\eta'=2$ is obtained from the 30 left-right sweeps as shown in Fig.~\ref{fig:test_mm_fp}.
The ground state wavefunction (GS) is the isoTNS obtained from DMRG\textsuperscript{2} and is also the initial state for the repeating MM sweep.
}
\end{figure}

We take the approximate fixed point wavefunction of $D'=\eta'=2$ in Fig.~\ref{fig:test_mm_fp} after 30 left-right sweeps and compute the connected correlation functions $\braket{\sigma^z_{i,5}\sigma^z_{5,5}}_{\textrm{c}}$ and $\braket{\sigma^x_{i,5}\sigma^x_{5,5}}_{\textrm{c}}$  along the horizontal line across the center of the lattice.
We plot the comparison with the result obtained with the ground state wavefunction in Fig.~\ref{fig:fp_gs_corr}.
The result suggest the correlation functions remain unchanged under multiple MMs.

We list the ground state energy for Kitaev honeycomb mode obtained through $\text{DMRG}^2$ with 2D isoTNSs in Table~\ref{tab:GS}.

\makeatletter\onecolumngrid@push\makeatother
\begin{table*}[h]
\caption{\label{tab:GS}%
Table for energy per site $E_0/N$ of Kitaev honeycomb model at isotropic point $J_x= J_y = J_z = 1$.
}
\begin{ruledtabular}
\begin{tabular}{cccccc}
System Size\footnote{$N = L_x \times L_y \times 2$} & \textrm{Exact} & \textrm{DMRG}& \textrm{DMRG} & \textrm{isoTNS} &  \textrm{isoTNS} \\
$(L_x, L_y)$ & &  $\chi = 512$ &  $\chi = 1024$ & $D=4 ,\eta=8$ & $D=6 ,\eta=12$\\
\hline
(5, 5)      & -0.71402401   & -0.71362916   & -0.71401082   & -0.70689939   & \\
(7, 7)      & -0.73416737   & -0.72593548   & -0.73039993   & -0.72696637   & -0.72987357 \\
(11, 11)    & -0.75303346   & -0.72405548   & -0.72808531   & -0.74448479   & -0.74682331 \\
\end{tabular}
\end{ruledtabular}
\end{table*}

\clearpage
\makeatletter\onecolumngrid@pop\makeatother

\bibliography{apssamp}%

\end{document}